
\magnification=\magstep1          \def\sp{\,\,\,} \overfullrule=0pt
  \def\c{\chi} 

\def\l{\Lambda}   \def\la{\lambda}    \def\Z{{\bf Z}}
\def\u{\tau}       \def\equi{\,{\buildrel \rm def \over =}\,}
\def\eg{{\it e.g.}$\sp$} \def\ie{{\it i.e.}$\sp$}
 \def\g{{\hat g}}

  \def\GW{[2]}  \def\RT{[15]} \def\COMM{[7]} \def\HET{[9]}
\def\ITZ{[6]} \def\CIZ{[5]} \def\RTW{[8]} \def\VER{[22]}
\def\BMW{[23]} \def\CR{[10]} \def\KAC{[4]} \def\MS{[11]}
\def\HK{[12]} \def\CS{[14]} \def\ROB{[24]} \def\KMPS{[13]}
\def\SCH{[16]} \def\VS{[21]} \def\ALZ{[20]} \def\GR{[25]}
\def\PM{[19]} \def\VT{[26]} \def\WAL{[27]} \def\NM{[18]}
\def\BI{[17]} \def\HAL{[3]} \def\WN{[1]} \def\WG{[1,2]}
\def\CM{[10,11]} \def\KK{[4,13]} \def\PF{[18,19]} \def\SM{[16,11]}

{ \nopagenumbers
\rightline{December, 1992}
\bigskip \bigskip
\centerline{{\bf The Classification of Affine}}
\bigskip
\centerline{{\bf SU(3) Modular Invariant Partition Functions}}
\bigskip \bigskip\bigskip
\centerline{{Terry Gannon}}
\centerline{{\it Mathematics Department, Carleton University}}
\centerline{{\it Ottawa, Ontario, Canada K1S 5B6}}\bigskip
\bigskip \bigskip \bigskip
A complete classification of the WZNW modular invariant partition functions is
known for
very few affine algebras and levels, the most significant being all levels of
SU(2), and level 1 of all simple algebras. In this paper we solve the
classification problem for SU(3) modular invariant partition functions.
Our approach will also be applicable to other affine Lie algebras, and we
include some preliminary work in that direction, including a sketch of a new
proof for SU(2).
\vfill \eject} \pageno=1

\centerline{{\bf 1. Introduction}} \bigskip

The classification of all rational conformal field theories is clearly a
desirable goal, but the scarcity of concrete results indicates its extreme
difficult, if not impossibility. This program includes the more manageable
but still very difficult
classification of all modular invariant partition functions for each
choice of affine algebra and level.

The partition function of a Wess-Zumino-Novikov-Witten conformal field theory
\WG{} associated with  affine Lie algebra \HAL{}
 $\g$ and level $k$ can be
 written in the following way:
$$Z=\sum N_{\lambda_L \lambda_R} \c_{ \lambda_L}^{k}
\,\c_{\lambda_R}^{k*}. \eqno(1.1)$$
$\c_{\lambda}^{k}$
 is the {\it normalized character} \KAC{} of the representation of
 $\g$ with (horizontal) highest weight $\lambda$ and level $k$; it is a
function of a complex vector $z$ and a complex number $\tau$.
The algebra $\g$ is the untwisted
affine extension $g^{(1)}$ of a simple Lie algebra $g$ (this extends in
the obvious way to semi-simple algebras).
The (finite) sum in eq.(1.1) is over  the horizontal highest weights
$ \lambda_{L},\la_R$ of level $k$.

There are three properties the sum in eq.(1.1) must satisfy in order
to be the partition function of a physically sensible conformal field theory:
\medskip
\item{(P1)} {\it modular invariance}. This is equivalent to
the two conditions:
$$\eqalignno{Z(z_Lz_R|\u+1)=&Z(z_Lz_R|\u), &(1.2a)\cr
\exp[-k\pi i(z_L^2/\u-z_R^{*2}/\u^*)]\,
Z(z_L/\u, z_R/\u|-1/\u)=&Z(z_L z_R|\u); &(1.2b)\cr}$$

\item{(P2)} {\it positivity} and {\it integrality}. The coefficients
$N_{ \lambda_L \lambda_R}$
in eq.(1.1) must be non-negative integers; and

\item{(P3)} {\it uniqueness of vacuum}. $\la=0$ is a possible
highest weight vector, for any $g$ and $k$. We must have $N_{00}=1$
(in the following sections we will change notations slightly, and this
will become $N_{\rho\rho}=1$).\medskip

We will call any modular invariant function $Z$ of the form (1.1),
an {\it invariant}. $Z$ will be called {\it positive} if
in addition each $N_{\la_L\la_R}\ge 0$, and {\it physical}
 if it satisfies (P1), (P2), and (P3).
Our task is to find all physical invariants corresponding
to each algebra $g$ and level $k$.

An invariant satisfying (P1), (P2) and (P3) is still not necessarily the
partition
function of a conformal field theory obeying duality and CPT-invariance. If it
is, we will call it {\it strongly physical}. These are the invariants of
interest to physics. We will discuss the additional properties
satisfied by strongly physical invariants (most importantly, that they
become automorphism invariants when written in terms of the characters of
their maximal chiral algebras) at the beginning of Sec.5.

 Much work has been done over the past few years on finding these physical
 invariants. But there has been comparatively little progress in the
 task of determining all physical invariants belonging to certain choices
of $g$ and $k$: all physical invariants for $g=A_1$ are known, for any
level $k$ \CIZ; all level 1 physical invariants have been found for simple
$g$ --- namely, $g=A_n$ \ITZ, and $g=B_n,$ $C_n$, $D_n$, $E_{6,7,8}$,
$F_4$, $G_2$ \COMM;
and all $A_2$ level $k$ ones are known when $k+3$ is prime \RTW.
Some work in classifying the {\it heterotic} physical invariants has also
been done \HET.

Unfortunately, enough simplifications apply to the level 1 cases,
and to the $A_1$ case, to make it unclear how to extend those arguments
to more general cases.
In this paper we will focus on the case $g=A_2$, although our primary interest
lies in developing tools applicable to other algebras (see Sec.6).
There are
several known physical invariants for $A_2$ \CM. These will be given in
eqs.(2.7). The question this paper addresses is the completeness
of this list. Two results in this direction are already known: the
list is complete for $k+3$ prime \RTW; the list is complete for $k\le 32$
\HK.

In Sec.2 we will introduce the notation and terminology used in the later
sections, and sketch the strategy taken. Sec.3 will find all
{\it permutation invariants} (see eq.(3.1)) of $A_2$, for each
level. In Sec.4 we find, for each $k$, a list of weights $\la$ for which
$N_{0\la}$ can be non-zero for some level $k$ physical invariant
$N$; this list shows, among other things, that the only $A_2$ physical
invariants for $k\equiv 2,4,7,8,10,11$ (mod 12) are permutation invariants.
Thus Secs.3 and 4 succeed in finding all $A_2$ physical invariants for
those levels. In Sec.5 we complete the classification for the remaining
levels (except for the levels 3,5,6,9,12,15 and 21, which we avoided because
of extra complications arising at those $k$),
but to do this we need to impose further physical conditions (namely, the
duality and CPT-invariance of the underlying conformal field theory) so that
the powerful analysis of
\MS{} can be applied --- we find all
$A_2$ strongly physical invariants for those levels. Together with \HK{} this
concludes the $A_2$ classification problem. In the final section we
investigate how well this approach extends to other algebras. The appendix
includes a detailed sketch of how this approach applies to $A_1$.

The key advantage the approach developed in this paper has over previous
approaches is that explicit construction of the commutant is avoided,
and positivity is imposed from the beginning. This significantly simplifies
the analysis required.

The only remaining question for the $A_2$
classification problem is to see if our proof, which found all {\it physical}
invariants for half the levels and all {\it strongly physical} ones for the
other half, can be strengthened so as to find all {\it physical} ones for all
levels --- although all assumptions we have imposed are physically valid,
it would be nice to reduce these to the smallest number possible.
A more interesting and important question is to find other algebras
which can be handled by analogous methods.

\bigskip \bigskip \centerline{{\bf 2. Terminology and sketch of proof}}
\bigskip

Before we begin the main body of this paper, it is necessary to introduce
some notation and terminology.
For a much more complete description of the rich theory of Ka{\v c}-Moody
algebras,
see \eg \KK. We will restrict attention here to the algebra $g=A_2$,
 but similar comments hold for the other algebras. The few facts about
lattices which we need are included in \eg \CS.

The root=coroot lattice of $g=A_2$ is also called $A_2$. Let
$\beta_1,\beta_2$ denote the fundamental weights of $A_2$, and write
$\rho=\beta_1+ \beta_2$; $\beta_1$ and $\beta_2$ span the {\it dual lattice}
$A_2^*$ of $A_2$.
Throughout this paper we will identify the weight $\la=m\beta_1+n\beta_2$ with
its Dynkin labels $(m,n)$.

An integrable irreducible  representation of the affine Lie algebra
$\g=A_2^{(1)}$ is given by a positive integer
$k$ (called the {\it level}) and a {\it highest weight} $\la\in A_2^*$.
The set of all possible highest weights corresponding to  level $k$
representations is
$$P_{+}^k\equi  \{m\beta_1+n\beta_2\,|\,m,n\in\Z,\sp
0\le m,n,\sp m+n\le k\}.\eqno(2.1a)$$
We will find it more convenient to use instead the related set
$$P^k=P^{k+3}_{++}\equi  \{m\beta_1+n\beta_2\,|\,m,n\in\Z,\sp
0< m,n,\sp m+n< k+3\}.\eqno(2.1b)$$
Clearly, $P^k=P_{+}^k+\rho$, and $\rho\in P^k$. {\it For the remainder of this
paper},
the character corresponding to the level $k$ representation with highest
weight $\la=m\beta_1+n\beta_2\in P_{+}^k$ will be denoted
$$\chi^k_{\la+\rho}=\chi_{m+1,n+1}^k.$$
The trivial representation of level $k$, which is given by highest weight
$\la=0$, corresponds then to the character $\chi_{\rho}^k=\chi_{11}^k$, and
(P3) becomes $N_{11,11}=1$.

Let $\hat{\alpha}_0,\hat{\alpha}_1,\hat{\alpha}_2$ be the simple roots of
$\hat{A_2}$. The 6 outer automorphisms of $\hat{A_2}$ are generated by
$h$ (order 2) and $\omega$ (order 3), where $h(\hat{\alpha}_0)=\hat{\alpha}_0,
h(\hat{\alpha}_1)=\hat{\alpha}_2,h(\hat{\alpha}_2)=\hat{\alpha}_1$, and
$\omega(\hat{\alpha}_0)=\hat{\alpha}_1,\omega(\hat{\alpha}_1)=\hat{\alpha}_2,
\omega(\hat{\alpha}_2)=\hat{\alpha}_0$. On the weights $(m,n)\in P^k$ these
become
$$\eqalignno{h(m,n)&=(n,m),&(2.2a)\cr \omega(m,n)&=(k+3-m-n,m).&(2.2b)\cr}$$
Note that $\omega^2(m,n)=(n,k+3-m-n)$. We will be encountering $h$ and
$\omega$ throughout the paper.

The {\it Weyl-Ka{\v c} character formula} gives us a convenient expression for
the character $\c^{k}_\la$:
$$\eqalignno{\c_\lambda^{k}(z,\u)=&{\sum_{w\in W} \epsilon(w)
\Theta\bigl({\la\over \sqrt{k+3}}+\sqrt{k+3}A_2\bigr)(\sqrt{k+3}w(z)|\u)
\over D(z|\u)},&(2.3a)\cr {\rm where}\sp D(z|\u)\equi &\sum_{w\in W}
\epsilon(w)\Theta\bigl( {\rho \over \sqrt{3}}+\sqrt{3}A_2\bigr)(\sqrt{3}
w(z)|\u), &(2.3b)\cr
{\rm and}\sp\Theta\bigl(v+\l\bigr)(z|\u)\equi &\sum_{x\in \l}
\exp[\pi i \u(x+v)^2+2\pi iz\cdot (x+v)].&(2.3c)\cr}$$
Here, $W$ is the 6 element Weyl group of $A_2$ and $\epsilon(w)=$ det $w\in
\{\pm 1\}$. The variable $\u\in {\bf C}$ satisfies Im $\u>0$, and $z=z_1\beta_1
+z_2\beta_2$ is a complex vector. Unlike much of the literature, we will
retain $z\ne 0$, so an invariant here will usually be different from its {\it
charge conjugate} (2.7$h$).

By the {\it commutant} $\Omega^k$ we mean the (complex)
space of all functions
$$Z(z_Lz_R|\u)=\sum_{\la,\la'\in P^k} N_{\la\la'} \c^k_\la(z_L,\u)\,
\c^k_{\la'}(z_R,\u)^*\eqno(2.4)$$
invariant under the modular group, \ie those $Z$ in (2.4) satisfying
(P1). It is not hard to show that two functions $Z$ and $Z'$ are
equal iff their {\it coefficient} (or {\it mass) matrices} $N$ and $N'$ are
equal; we will use
the invariant $Z$ interchangeably with its matrix $N$.

The functions $\c^k_\la$ behave quite nicely under the modular
transformations $\u\rightarrow \u+1$ and $\u\rightarrow -1/\u$:
$$\eqalignno{\c_\la^k(z,\u+1)=&\sum_{\la'\in P^k}\big(T^{(k)}
\big)_{\la\la'}\c_{\la'}^{k}(z,\u),\sp \sp{\rm where}&(2.5a)\cr
\big(T^{(k)}\big)_{\la\la'}=&\exp[\pi i{ \la^2\over k+3}-\pi i
{2\over 3}] \,\delta_{\la\la'} &(2.5b)\cr
=&e_k(-m^2-mn-n^2+k+ 3)\,\delta_{m,m'}\,\delta_{n,n'};&(2.5c)\cr
\c_\la^{k}(z/\u,-1/\u)=&\exp[k\pi iz^2/\u]\,\sum_{\la'\in P^k}\big(S^{(k)}
\big)_{\la\la'}\c_{\la'}^{k}(z,\u),\sp \sp{\rm where}&(2.5d)\cr
\big(S^{(k)}\big)_{\la\la'}=&{i^{3}\over (k+3)\sqrt{3}}
\sum_{w\in W}\epsilon(w)\exp[-2\pi i {w(\la')\cdot\la\over k+3}] &(2.5e)\cr
=&{-i\over \sqrt{3}(k+3)}\bigr\{ e_k(2mm'+mn'+nm'+2nn')+e_k(-mm'-2mn'-nn'+nm')
&\cr & +e_k(-mm'+mn'-2nm'-nn')-e_k(-2mn'-mm'-nn'-2nm')&\cr &
-e_k(2mm'+mn'+nm'-nn')-e_k(-mm'+mn'+nm'+2nn')\},&(2.5f)\cr}$$
where in (2.5$c,f$) we have $\la=m\beta_1+n\beta_2$,
$\la'=m'\beta_1+n'\beta_2$ and the function $e_k$ is defined by
$e_k(x)\equi \exp[{-2\pi i x\over 3(k+3)}]$. The matrices $T^{(k)}$ and
$S^{(k)}$ are unitary and symmetric.

Note that $Z=\sum N_{\la\la'}\c^k_\la\c^{k*}_{\la'}\in \Omega^k$ iff
both $$\eqalignno{\bigl(T^{(k)}\bigr)^{\dag}\,N\,\bigl(T^{(k)}\bigr)=&
N,&(2.6a)\cr
\bigl(S^{(k)}\bigr)^{\dag}\,N\,\bigl(S^{(k)}\bigr)=&
N.&(2.6b)\cr}$$

Recall the outer automorphisms $h$ and $\omega$ given in (2.2).
The known physical invariants of $A_2$ are:
$$\eqalignno{{\cal A}_k\equi & \sum_{\la\in P^k} |\c_\la^k|^2,&(2.7a)\cr
{\cal D}_k\equi & \sum_{(m,n)\in P^k}\chi^k_{m,n}\,\chi^{k*}_{\omega^{k(m-n)}
(m,n)},\quad {\rm \sp for \sp}k\not\equiv 0 \sp{\rm (mod\sp 3) \sp and
\sp}k\ge 4;&(2.7b)\cr
{\cal D}_k\equi & {1\over 3}
\sum_{{(m,n)\in P^k \atop m\equiv n \sp({\rm mod}\sp 3)}}
\,|\chi^k_{m,n}+\chi_{\omega(m,n)}^k+\chi^k_{\omega^2(m,n)}|^2,
{\rm \sp for \sp}k\equiv 0 \sp{\rm (mod\sp 3)};&(2.7c)\cr
{\cal E}_5\equi & |\c^5_{1,1}+\c^5_{3,3}|^2+|\c^5_{1,3}+\c^5_{4,3}|^2+
|\c^5_{3,1}+\c^5_{3,4}|^2&\cr & +|\c^5_{3,2}+\c^5_{1,6}|^2+|\c^5_{4,1}+
\c^5_{1,4}|^2+|\c^5_{2,3}+\c^5_{6,1}|^2;&(2.7d)\cr
{\cal E}_9^{(1)}\equi &|\c^9_{1,1}+\c^9_{1,10}+\c^9_{10,1}+\c^9_{5,5}+
\c^9_{5,2}+\c^9_{2,5}|^2+2|\c^9_{3,3}+\c^9_{3,6}+\c^9_{6,3}|^2;&(2.7e)
\cr {\cal E}_9^{(2)}\equi &|\c^9_{1,1}+\c^9_{10,1}+\c^9_{1,10}|^2+|\c^9_{3,3}+
\c^9_{3,6}+\c^9_{6,3}|^2+2|\c^9_{4,4}|^2& \cr &+|\c^9_{1,4}+\c^9_{7,1}+
\c^9_{4,7}|^2+|\c^9_{4,1}+\c^9_{1,7}+\c^9_{7,4}|^2+|\c^9_{5,5}+\c^9_{5,2}+
\c^9_{2,5}|^2&\cr &+(\c^9_{2,2}+\c^9_{2,8}+\c^9_{8,2})\c^{9*}_{4,4}+
\c^9_{4,4}(\c^{9*}_{2,2}+\c^{9*}_{2,8}+\c^{9*}_{8,2});&(2.7f)\cr
{\cal E}_{21}\equi & |\c^{21}_{1,1}+\c^{21}_{5,5}+\c^{21}_{7,7}+
\c^{21}_{11,11}+\c^{21}_{22,1}+\c^{21}_{1,22}&\cr &+\c^{21}_{14,5}+
\c^{21}_{5,14}+\c^{21}_{11,2}+\c^{21}_{2,11}+\c^{21}_{10,7}+\c^{21}_{7,10}
|^2&\cr &+|\c^{21}_{16,7}+\c^{21}_{7,16}+\c^{21}_{16,1}+\c^{21}_{1,16}+
\c^{21}_{11,8}+\c^{21}_{8,11} &\cr &+\c^{21}_{11,5}+\c^{21}_{5,11}+
\c^{21}_{8,5}+\c^{21}_{5,8}+\c^{21}_{7,1}+\c^{21}_{1,7}|^2;&(2.7g)\cr}$$
together with their {\it conjugations} $Z^c$ under $h$, defined by:
$$Z^c=\sum_{\la,\la'\in P^k}N_{\la,h(\la')}\c_\la^k \c^{k*}_{\la'}=
\sum_{m,n,m',n'}N_{mn,n'm'}\c_{mn}^k\c_{m'n'}^{k*},\eqno(2.7h)$$
where $Z$ is given by (2.4).
Note that ${\cal D}_3={\cal D}_3^c$, ${\cal D}_6={\cal D}_6^c$,
${\cal E}_9^{(1)}={\cal E}_9^{(1)c}$, and ${\cal E}_{21}={\cal E}_{21}^c$.
In the case of restricted characters $\c(0,\u)$, $Z=Z^c$.

Our goal is to prove that this list is complete: in particular we will prove

\bigskip\noindent{\bf Theorem 1(a)}:\quad For $k\equiv 2,4,7,8,10,11$ (mod 12),
and $k=1$, the set of all {\it physical invariants}
for $A_2$ is given by eqs.(2.7);

{\bf 1(b)}:\quad for $k\equiv 0,1,3,5,6,9$ (mod 12), $k\ne 3,5,6,9,12,15,21$,
the set of all {\it strongly physical invariants} for $A_2$ is given by
eqs.(2.7).\bigskip

(The terms {\it physical invariant} and {\it strongly physical invariant}
are defined in Sec.1.)

Two partial results are already known. In \RTW{} this theorem is proven for
$k+3$ prime. They accomplish this by very explicitly computing a basis for
the commutant, then finding all the positive invariants, and lastly
imposing the uniqueness condition $N_{11,11}=1$. Unfortunately this
explicitness
makes it very difficult to apply their approach to more general $k$. A
second partial result is the computer search in \HK. Using the
Roberts-Terao-Warner lattice method \RT, it finds a basis for the
commutant for a given $k$, and then imposes positivity and uniqueness of the
vacuum. The proof given in \COMM{} that lattice partition functions span
the commutant guarantees the completeness of this search.
In this way it has verified that the list in eqs.(2.7) is complete for
all $k\le 32$
(it also applies this technique to the three other rank 2 algebras). This
program thus fills in all of the holes of Thm.1(b).

The approach taken here is somewhat different.

Call an invariant $\rho${\it -decoupled} if $N_{\rho,\la}
=N_{\la,\rho}=0$ for all $\la\neq \rho$. Hence such an invariant can be
written in the form
$$Z=a|\c_\rho^k|^2+\sum_{\la,\la'\ne \rho}N_{\la,\la'}\c_\la^k\c_{\la'}^{k*}$$
For example, the only $\rho$-decoupled invariants in eqs.(2.7) are
(2.7$a,b$) and their conjugates. A valuable observation was made in
\COMM{} (see also \SM):

\medskip \noindent{\bf Lemma 1}:\quad A $\rho$-decoupled physical invariant
is a {\it permutation invariant} (defined in eq.(3.1) below).\medskip

All permutation invariants are found in the following section. In Sec.4 we
proceed to show that for some levels, any physical invariant must be a
permutation invariant, thus proving Thm.1 for those levels.

A second observation made in \COMM{} connects more directly with the lattice
method of \RT. First note the following:

It is proven in \KAC{} that for any $\la\in A_2^{*}$, either
$$\sum_{w\in W}\epsilon(w)\,\Theta\left({w(\la)\over \sqrt{k+3}}+\sqrt{k+3}
A_2\right)/D=0 \eqno(2.8a)$$
holds identically (where $\Theta$ and $D$ are defined in eqs.(2.3$c,b$),
respectively), or there exists a $w'\in W$
and a $\la'\in P^k$ such that $\la\equiv w'(\la')$ (mod $(k+3)A_2)$, and hence
$$\sum_w\epsilon(w)\,\Theta\left({w(\la)\over \sqrt{k+3}}+\sqrt{k+3}A_2
\right)/D=\epsilon(w')\,\c^k_{\la'}. \eqno(2.8b)$$
By the {\it parity} $\epsilon(\la)$ of $\la$ we mean $\epsilon(\la)=0$ if
(2.8$a$) holds, and $\epsilon(\la)= \epsilon(w')$ if (2.8$b$) does. When
$\epsilon(\la)\ne 0$, let $[\la]_k$ denote the (unique) weight $\la'$
in (2.8$b$).

Now choose any $\la_L,\la_R\in P^k$. We showed in \COMM, {\it for each $\ell$
relatively prime to} $3(k+3)$, that $\epsilon(\ell\la_L)
\epsilon(\ell\la_R)\ne 0$,
and that for any level $k$ invariant $Z$ in (2.4),
$$N_{\la_L\la_R}=\epsilon(\ell\la_L)\,\epsilon(\ell\la_R)\,N_{[\ell\la_L]_k
[\ell\la_R]_k}.\eqno(2.9)$$
We did this by first showing it for lattice partition functions, where it
is obvious, and then referring to the result that lattice partition functions
span the commutant. A similar derivation of (2.9) can be made using the
construction in \BI{} (and generalized in \COMM{}) of the {\it Weyl-unfolded
commutant}.

``$\ell$ relatively prime to $3(k+3)$'' is equivalent here (and in Lemma 2
below) to the statement
``$\ell$ relatively prime to the order $L$ of the vector $(\la_L;\la_R)$
with respect to the lattice $((k+3)A_2;(k+3)A_2)$'' --- indeed that is how
(2.9) is expressed in \COMM.
Examples of (2.9) for $k=5$ and $k=9$ are given in \COMM. Of course,
it also holds for all other algebras.
Eq.(2.9) (as well as Lemma 2 below) is used in \HK{} to eliminate `redundant'
coefficients $N_{\la\la'}$, and hence moderate memory problems.
Its main value for our purpose lies in its trivial consequence:

\medskip\noindent{\bf Lemma 2:} \quad Let $\la,\la'\in P^k$.
 If some $\ell$ relatively
 prime to $3(k+3)$ satisfies $\epsilon(\ell \la)\epsilon(\ell\la')=-1$, then
$N_{\la,\la'}=0$ for any {\it positive invariant} $N$.\medskip

The analogue of Lemma 2 holds for all algebras. Lemma 2 constitutes an
extremely strong constraint on which $\la,\la'\in P^k$ may {\it couple}
--- \ie have $N_{\la,\la'}\ne 0$ --- in some positive invariant $N$.
It hints that  the space $\Omega^k_+$ spanned by the positive
invariants of level $k$ may have much smaller dimension than the full commutant
$\Omega^k$ and so may be a much more convenient
space to work with. Indeed, although the dimension of the commutant
$\Omega^k$ goes to infinity with $k$, dim $\Omega_+^k=4$ for many $k$ \HK. Our
approach
involves using Lemma 2 to keep our analysis restricted as much as possible
to the space $\Omega^k_+$, instead of $\Omega^k$.

A final tool that we will mention here also
holds for any positive invariant
of any algebra and level, and exploits the fact that the product $NN'$ of two
invariants is also an invariant (this can be read off from eqs.(2.6)).
It is proved using the Perron-Frobenius theory of
non-negative matrices \PF, and can be thought of as a generalization of
Thm.4 in \COMM. It will be used in Sec.5 to significantly restrict the
possibilities for the coefficient matrix $N$ of physical invariants.

Any matrix $M$ can be written as a direct sum $\oplus_i M_i$ of {\it
indecomposable} blocks $M_i$.
By a {\it non-negative matrix} we mean a square matrix $M$ with non-negative
real entries. Any such matrix has a non-negative real
eigenvalue $r=r(M)$ with the
property that $r\ge |s|$ for all other (possibly complex) eigenvalues $s$
of $M$. The number $r(M)$ has many nice properties, for example:
$${\rm min}_i \sum_j M_{ij}\le r(M)\le {\rm max}_i \sum_j M_{ij},\eqno(2.10a)
$$
and if $M$ is indecomposable, either equality holds iff each row sum $\sum_j
M_{ij}$ is equal;
and $${\rm max}_i M_{ii}\le r(M),\eqno(2.10b)$$
and if $M$ is indecomposable and symmetric, equality happens in (2.10$b$)
iff $M$
is a $1\times 1$ matrix $M=(M_{11})$. Also, there is an eigenvector $v$
with eigenvalue $r$ whose components $v_i$ are all non-negative reals.

\medskip\noindent{\bf Lemma 3}:\quad Let $Z=\sum N_{\la\la'}\chi_{\la}
\chi_{\la'}$ be a positive invariant, for any algebra and any level.
Write $N$ as a direct sum of indecomposable blocks
$$N=\bigoplus_{\ell=0}^L N_\ell=\left(
\matrix{N_0&0&\cdots&0\cr 0&N_1&\cdots&0\cr
\vdots&&&\vdots\cr 0&0&\cdots&N_L\cr}\right),\eqno(2.10c)$$
where $N_0$ is the block `containing' $N_{\rho\rho}$.
Then $r(N_\ell)\le r(N_0)$ for all $\ell$. If in addition $N$ is a
symmetric matrix, and if for all $\ell$ with $r(N_\ell)=r(N_0)$ we have
$(N_\ell)^2=c_\ell N_\ell$ for some constant $c_\ell$,
then for all $m$, either $r(N_m)=r(N_0)$ or $N_m=(0)$.\medskip

\noindent{\it Proof}\quad
Suppose $r(N_{\ell_0})>r(N_0)$ for some $\ell_0$, and
choose any $r$ satisfying $r(N_0)<r<r(N_{\ell_0})$. Consider
the limit as $n\rightarrow \infty$ of each $({1\over r}N_\ell)^n$.
It is easy to show (\eg using Jordan blocks) that
if all eigenvalues $\la$ of a  matrix $M$ have norm $|\la|<1$, then the
limit of $M^n$ is the 0-matrix. In particular, the limit of $({1\over r}N_0)^n$
will be 0. What happens to $({1\over r}N_{\ell_0})^n$?

Let $v$ be an eigenvector of $N_{\ell_0}$ with eigenvalue
$r_{\ell_0}=r(N_{\ell_0})$, whose components are non-negative reals. Then
$({1\over r}N_{\ell_0})^n v=(r_{\ell_0}/r)^nv$. By positivity, this
implies that $({1\over r}N_{\ell_0})^n$ will have some arbitrarily large
components as $n$ increases.

The matrix $({1\over r}N)^n$ will correspond to a  positive invariant, for
each $n$, and will be the direct sum of the blocks $({1\over r}N_\ell)^n$.
Taking $n$ sufficiently large, eq.(5.2) of \COMM{} can now be used to give
us a contradiction.

Thus $r(N_\ell)\le r(N_0)$.

If $N_\ell^2=m_\ell N_\ell$, then by the above argument $r(N_\ell)=m_\ell$.
The remainder of the proof is as in Thm.4 of \COMM.
\qquad QED\medskip

The conditions in the last sentence of the lemma can be weakened somewhat,
but this is all that we will need in this paper.
A commonly occurring example of a matrix $M$ with the property $M^2=mM$ is
the $n\times n$ matrix
$$M_{n,\ell}=\left(\matrix{\ell&\ell&\cdots&\ell\cr \vdots&\vdots&&\vdots
\cr\ell&\ell&\cdots&\ell\cr}\right).\eqno(2.10d)$$
Here, $r(M_{n,\ell})=m=\ell n$.

The strategy adopted in this paper is three-fold.

\item{{\it Sec.{} 3}} Find all permutation invariants for each level $k$.
We accomplish this by repeatedly exploiting the facts that this permutation
must be a symmetry of both $S^{(k)}_{\la\mu}$ and the fusion rules
$N^{(k)}_{\la\mu\nu}$.

\item{{\it Sec.{} 4}} For each $k$, use Lemma 2 to find all weights
$\lambda\in P^k$ which can couple to $\rho$ in some
positive invariant $N$. The argument is elementary but tedious and involves
investigating several cases. There are surprisingly few such $\la$; the
results are compiled in Lemma 4. There will always be at least one such
weight, namely $\rho$ itself. When this is the only one, then Lemma 1 tells us
that any physical invariant of that level must necessarily be a permutation
invariant, and so must be on the list found in Sec.{} 3.

\item{{\it Sec.{} 5}} The remaining levels, which have nontrivial
$\rho$-couplings, must now be considered. To do them,
we use \MS, together with Lemma 3, to write down the characters of all
possible maximal extensions of $\hat{A_2}$ consistent with Lemma 4; if there
are any such extensions,
we then find their symmetries by mimicking the argument of Sec.3.

In this paper we only make use of Lemma 2 for $\lambda'=\rho$. It is quite
possible that applying it to other weights will permit us to avoid using
\MS{} in Sec.5, and so could yield a classification proof for those levels
which assumes only (P1), (P2), (P3), instead of exploiting in addition the
existence and properties of the maximally extended chiral algebras of the
theory. A more interesting possibility is to exploit more of the rich
algebraic structure of $\Omega^k$.

\bigskip \bigskip \centerline{{\bf 3. The permutation invariants}}
\bigskip

By a {\it permutation invariant} (sometimes called an {\it automorphism
invariant}) we mean an invariant of the form
$$\eqalignno{Z=&\sum_{\la\in P^k} \c_\la \, \c_{\sigma \la}^*, &(3.1a)\cr
i.e.\sp N_{\la\la'}=&N^\sigma_{\la\la'}\equi
\delta_{\la',\sigma\la}&(3.1b)\cr}$$
for some permutation $\sigma$ of $P^k$. In this section we will find all
$A_2$ permutation invariants, for each $k$. In particular, we will prove the
following theorem:

\medskip\noindent{{\bf Theorem 2}}: \quad The only level $k$ permutation
invariants for $A_2$ are ${\cal A}_k$, ${\cal A}_k^c$ for $k\equiv 0$ (mod 3),
and ${\cal A}_k$, ${\cal A}^c_k$, ${\cal D}_k$, ${\cal D}_k^c$ for
$k\not\equiv 0$ (mod 3). \medskip

Many permutation invariants, for each algebra, have been constructed (see
\eg \ALZ), but their methods cannot claim to find them all. For example,
the $k=4$ $G_2$ and $k=3$ $F_4$ exceptional permutation invariants found
in a computer search in \VS{} were missed by \ALZ, and also cannot be
obtained using simple currents \VS. Until now, only for $A_1$ \CIZ{} have
all permutation invariants been classified; recently \GR{} all those for
$g=A_1\oplus\cdots\oplus A_1$ have also been found.

Throughout this section let $k'=k+3$, and assume $N^\sigma$ is a permutation
invariant.

That the matrix $N^\sigma$ in (3.1$b$) must commute with $S^{(k)}$
and $T^{(k)}$ (see (2.6)) is equivalent to
$$\eqalignno{S^{(k)}_{\la\la'}&=S^{(k)}_{\sigma \la,\sigma \la'},&(3.2a)\cr
T^{(k)}_{\la\la'}&=T^{(k)}_{\sigma \la,\sigma\la'},&(3.2b)\cr}$$
for all $\la,\la'\in P^k$. Note that (2.5$c$) tells us that (3.2$b$) is
equivalent to the condition that
$$m^2+mn+n^2\equiv m'{}^2+m'n'+n'{}^2 \sp({\rm mod}\sp 3k')\eqno(3.2c),$$
for all $(m,n)\in P^k$, where $\sigma(m,n)=(m',n')$.

It can be shown (Thm.3 in \COMM) that any permutation invariant must be
physical, so
$$\sigma(1,1)=(1,1).\eqno(3.3a)$$
Also, we know $(S^{(k)})^2=C^{(k)}$, the {\it charge conjugation matrix}
defined
by $C^{(k)}_{mn,m'n'}=\delta_{m,n'}\delta_{n,m'}$, so $N^\sigma$ must
commute with $C^{(k)}$. This means
$$\sigma(m,n)=(m',n')\sp {\rm iff}\sp \sigma(n,m)=(n',m').\eqno(3.3b)$$

Verlinde's formula \VER{} gives us a relation between the fusion coefficients
$N^{(k)}_{\la\mu\nu}$ and the $S^{(k)}$ matrix:
$$N^{(k)}_{\la\mu\nu}=\sum_{\la'\in P^k} {S^{(k)}_{\la\la'}S^{(k)}_{\mu\la'}
S^{(k)}_{\nu\la'}\over S^{(k)}_{\rho\la'}}.\eqno(3.4a)$$
Therefore (3.2$a$) tells us that
$$N^{(k)}_{\la\mu\nu}=N^{(k)}_{\sigma\la,\sigma\mu,\sigma\nu}.\eqno(3.4b)$$
(3.4$b$) is useful to us, because these fusion coefficients have been
computed for $A_2$ \BMW. The formula will be given in the following paragraph.

Write $\la=\la_1\beta_1+\la_2\beta_2$, $\mu=\mu_1\beta_1+\mu_2\beta_2$,
$\nu=\nu_1\beta_1+\nu_2\beta_2$. Define
$$\eqalign{A&={1\over 3}[2(\la_1+\mu_1+\nu_1)+\la_2+\mu_2+\nu_2],\cr
 B&={1\over 3}[\la_1+\mu_1+\nu_1+2(\la_2+\mu_2+\nu_2)],\cr
k_{min}&={\rm max}\bigl\{\la_1+\la_2,\mu_1+\mu_2,\nu_1+\nu_2,A-
{\rm min}\{\la_1,\mu_1,\nu_1\},B-{\rm min}\{\la_2,\mu_2,\nu_2\}\bigr\},\cr
k_{max}&={\rm min}\{A,B\},\cr
\delta&=\left\{\matrix{1&{\rm if}\sp k_{max}>k_{min} \sp {\rm and}\sp
 A,B\in \Z\cr 0&{\rm otherwise}\cr} \right. .\cr}$$
Then \BMW{} says (changing their notation slightly and recalling that $k'
=k+3$)
$$N^{(k)}_{\la\mu\nu}=\left\{\matrix{0&{\rm if}\sp k'\le k_{min} \sp {\rm or}
\sp \delta=0\cr k'-k_{min}&{\rm if}\sp k_{min}\le k'\le k_{max}\sp {\rm and}
\sp \delta=1 \cr k_{max}-k_{min} & {\rm if} \sp k'\ge k_{max}\sp {\rm and}
\sp\delta=1\cr} \right. .\eqno(3.5)$$

The first step in the proof of Thm.2 is to show that ``point-wise'' $\sigma$
acts like an outer automorphism:

\medskip\noindent{\bf Claim}: \quad $\sigma(m,n)\in\{(m,n),(n,m),(m,k'-m-n),
(n,k'-m-n),(k'-m-n,m),(k'-m-n,n)\}$.
\bigskip\noindent{\it Proof} \quad Take $\la=\mu=\nu=m\beta_1+n\beta_2$. Then
(3.5) becomes
$$N^{(k)}_{\la\la\la}=\left\{\matrix{{\rm min}\{m,n\} &{\rm if}\sp
k'\ge m+n+{\rm min}\{m,n\}\cr k'-m-n &{\rm otherwise}\cr} \right. .
\eqno(3.6a)$$
Define ${\cal S}^k_a=\bigl\{(m,n)\in P^k\mid m=a$ or $n=a\bigr\}$,
$\tilde{\cal S}^k_b=\bigl\{(m,n)\in P^k\mid m+n=b\bigr\}$.
Eqs.(3.4$b$) and (3.6$a$) now imply
$$\sigma(m,n)\in {\cal S}^k_m\cup{\cal S}^k_n\cup {\cal S}^k_{k'-m-n}
\cup\tilde{\cal S}^k_{k'-m}\cup
\tilde{\cal S}^k_{k'-n}\cup \tilde{\cal S}^k_{m+n}. \eqno(3.6b)$$

Now let us ask the question: when can $\sigma(m,n)=(m,n')$? Eqs.(3.2$a$) and
(3.3$a$) would then imply $S^{(k)}_{mn,11}=S^{(k)}_{mn',11}$. Eq.(2.5$f$)
reduces this to
$$\sin({2\pi n\over k'})-\sin({2\pi (n+m)\over k'})=\sin({2\pi n'\over k'})
-\sin({2\pi (m+n')\over k'}).\eqno(3.7a)$$
Define $f_\alpha(x)=\sin(x)-\sin(x+\alpha)$. We are interested in finding all
solutions $f_\alpha(x)=f_\alpha(y)$, where $x,y,\alpha >0$ and $x+\alpha,y+
\alpha<2\pi$. Note that the derivative $f_\alpha'(x)$ is positive for
$x\in (-\alpha/2,\pi-\alpha/2)$, and negative for $x\in (\pi-\alpha/2,2\pi-
\alpha/2)$. Also, $f_\alpha$ is symmetric about its local maxima:
$f_\alpha(x+\pi-\alpha/2)=f_\alpha(-x+\pi-\alpha/2)$. What these facts mean
is that, in the interval $x,y\in(0,2\pi-\alpha)$, $f_\alpha(x)=f_\alpha(y)$
has the two solutions $x=y$ and $y=-x+2\pi-\alpha$. Hence the only possible
solutions to (3.7$a$) are
$$n'=n\sp {\rm and} \sp n'=k'-m-n.\eqno(3.7b)$$

The identical calculation and conclusion holds for $\sigma(m,n)=(n',m)$. Thus
$$\sigma(m,n)\in {\cal S}^k_m\sp \Rightarrow \sp \sigma(m,n)\in \bigl\{(m,n),
(m,k'-m-n),(n,m),(k'-m-n,m)\bigr\}. \eqno(3.7c)$$
The remaining  five possibilities in (3.6$b$) reduce to identical arguments.
\qquad QED to claim \medskip

The claim, together with (3.2$c$), tells us that the only possibilities for
$\sigma(1,2)$ are:
$$\eqalignno{\sigma(1,2)\in &\bigl\{(1,2),(2,1)\bigr\}\sp {\rm if}\sp k\equiv
0\sp{\rm (mod}\sp 3),&(3.8a)\cr
\sigma(1,2)\in &\bigl\{(1,2),(2,1),(2,k),(k,2)\bigr\}\sp {\rm if}\sp k\equiv
1\sp{\rm (mod}\sp 3),&(3.8b)\cr
\sigma(1,2)\in &\bigl\{(1,2),(2,1),(k,1),(1,k)\bigr\}\sp {\rm if}\sp k\equiv
2\sp{\rm (mod}\sp 3).&(3.8c)\cr}$$
Note that the possibilities for $k\equiv 0$ (mod 3) are realized by
${\cal A}_k$ and ${\cal A}_k^c$, respectively, and for $k\equiv \pm 1$ (mod 3)
by ${\cal A}_k$, ${\cal A}_k^c$, ${\cal D}_k$ and ${\cal D}_k^c$, respectively.
Since the (matrix) product of two permutation invariants is another permutation
invariant, to prove Thm.2 for each $k$ it suffices to show that the only
permutation invariant satisfying $\sigma(1,2)=(1,2)$ is ${\cal A}_k$.

Suppose for contradiction that $\sigma(1,2)=(1,2)$, but $\sigma(1,a)=(a,1)$
for some $2\le a\le k+1$. Then $S^{(k)}_{12,1a}=S^{(k)}_{12,a1}$, \ie
$$c_k(5a+4)+c_k(a+5)+c_k(4a-1)=c_k(4a+5)+c_k(5a+1)+c_k(a-4),\eqno(3.9a)$$
where $c_k(x)=\cos(2\pi {x\over 3k'})$. Making the substitution $b=a+
{1\over 2}$, we would like to show that
$$p(b,k)\equi c_k(5b+{3\over 2})+c_k(b+{9\over 2})+c_k(4b-3)-c_k(4b+3)-
c_k(5b-{3\over 2})-c_k(b-{9\over 2})\eqno(3.9b)$$
does not vanish at $b={5\over 2},{7\over 2},\ldots,k+{3\over 2}$.

Using the obvious trigonometric identities, we can rewrite $p(b,k)$ as a
polynomial in $c_k(b)$ and $s_k(b)=\sin(2\pi {b\over 3k'})$ --- in particular,
$p(b,k)=p^{(k)}_5\bigl(c_k(b)\bigr)+s_k(b)\cdot p^{(k)}_4\bigl(c_k(b)\bigr)$,
where $p^{(k)}_5$ and $p^{(k)}_4$ are, respectively, degree 5 and 4
polynomials. Note from (3.9$b$) that $p(-b,k)=-p(b,k)$, so $p^{(k)}_5$ must
be identically zero, and
$$p(b,k)=s_k(b)\cdot p^{(k)}_4\bigl(c_k(b)\bigr).\eqno(3.9c)$$
We are interested in the roots of this function, in the range $b\in(0,{3\over
2}k')$. Since $s_k(b)$ does not vanish, and $c_k$ is one-to-one, for those $b$,
for fixed $k$ there can be at most 4 zeros for $p^{(k)}_4$ and hence
$p(b,k)$ in that range. But $b={1\over 2},{3\over 2},k+{5\over 2},k+{7\over 2}$
are 4 distinct zeros for $p(b,k)$. Therefore they are the {\it only} zeros
in the range $b\in (0,{3\over 2}k')$, and so $p(b,k)$ cannot vanish at
$b={5\over 2},{7\over 2},\ldots,k+{3\over 2}$. This means that we cannot
have both $\sigma(1,2)=(1,2)$ and $\sigma(1,a)=(a,1)$, for any $a=2,3,\ldots,
k+1$.

The other four possibilities $\sigma(1,a)=(1,k'-1-a),(a,k'-1-a),(k'-1-a,1),$
and $(k'-1-a,a)$ all succumb to similar reasoning. Thus we have shown:
$$\sigma(1,2)=(1,2)\sp\Rightarrow\sp\sigma(1,a)=(1,a)\sp \forall(1,a)\in P^k.
\eqno(3.10a)$$

Remember, to prove Thm.2 it suffices to show $\sigma(1,2)=(1,2)$ implies
$\sigma(a,b)=(a,b)$ $\forall (a,b)\in P^k$. Suppose instead $\sigma(1,2)=(1,2)$
but $\sigma(a,b)=(b,a)$. Take $\la=(a,b)$, $\mu=(b,1)$, $\nu=(1,a)$ and
$\la'=(b,a)$. Then $N^{(k)}_{\la\mu\nu}=N^{(k)}_{\la'\mu\nu}$, by $(3.10a)$,
$(3.3b$) and $(3.4b$). But eq.(3.5) tells us $N^{(k)}_{\la\mu\nu}=1$, while
$N^{(k)}_{\la'\mu\nu}=0$ unless $a=b$.

Similar calculations show $\sigma(a,b)=(a,k'-a-b)$ only when $b=k'-a-b$,
and $\sigma(a,b)=(k'-a-b,b)$ only when $a=k'-a-b$. The remaining two
anomolous possibilities are slightly more difficult: $\sigma(a,b)=(b,k'-a-b)$
only when $3b=k'$, $b\le a$; and $\sigma(a,b)=(k'-a-b,a)$ only when $3a=k'$,
$a\le b$.

Now, if $a>b=k'/3$ and $\sigma(a,b)=(b,k'-a-b)$, then $\sigma$ being a
permutation implies $\sigma(b,k'-a-b)\ne(b,k'-a-b)$, so from the above
paragraph we must have either $3(k'-a-b)=k'$ or $b\le k'-a-b$, \ie either
$a=b=k'/3$ or $a\le b$ --- a contradiction.
Therefore the only way for either remaining anomolous possibility to be
realized is if $a=b=k'/3$.

Thus, we have shown
$$\sigma(1,2)=(1,2)\sp\Rightarrow\sp \sigma(a,b)=(a,b)\sp \forall(a,b)\in P^k,
\eqno(3.10b)$$
\ie that the only permutation invariant with $\sigma(1,2)=(1,2)$ is the
identity, which concludes the proof of Thm.2.

\bigskip \bigskip \centerline{{\bf 4. The $\rho$-coupling lemma}}
\bigskip

Again write $k'=k+3$.
Let ${\cal R}^k$ be the set of all $\la\in P^k$ such that there exists
a positive invariant with $N_{\rho,\la}\neq 0$.
For example, the known $A_2$ physical invariants (2.7) tell us that
${\cal R}^5\supseteq \{(1,1),(3,3)\}$, ${\cal R}^6\supseteq \{(1,1),(7,1),
(1,7)\}$ and ${\cal R}^7\supseteq\{(1,1)\}$.
If ${\cal R}^k=\{\rho\}$ then by Lemma 1 any level $k$ physical invariant
will be a permutation invariant, and will be listed in Thm.2.

Let $\la=a\beta_1+b\beta_2\in {\cal R}^k$. Then it must satisfy
$\la^2\equiv \rho^2$ (mod $2k'$), \ie
$$a^2+ab+b^2 \equiv  3 \sp ({\rm mod\sp} 3k').\eqno(4.1a)$$

It is easy to see from that equation that any $\la\in {\cal R}^k$ must
have order $k'$ with respect to $k'A_2$, and hence the vector $(\rho;\la)$
has order $L=k'$ with respect to $(k'A_2;k'A_2)$. Now, investigating
the behavior of the Weyl group $W$ on $A_2^*$ allows a simple formula for
the parity $\epsilon(\mu)$ (see (2.8$a$)) of an arbitrary vector
$\mu=c\beta_1+d\beta_2$:
for any real number $x$ define by $\{x\}$ the unique number congruent to
$x$ (mod $k'$) satisfying $0\le \{x\}<k'$, then
$$\epsilon(\mu)=\left\{ \matrix{0 &{\rm if\sp} \{c\},\,\{d\}\sp {\rm or}\sp
\{c+d\}=0\cr
+1&{\rm if}\sp \{c\}+\{d\}<k'\sp{\rm and}\sp\{c\},\{d\},\{c+d\}>0\cr
-1&{\rm if}\sp \{c\}+\{d\}>k'\sp{\rm and}\sp\{c\},\{d\},\{c+d\}>0\cr}
\right.\eqno(4.1b)$$
Then Lemma 2 implies that:
$$\eqalignno{0<\{\ell\} <k'/2,&\sp\ell\sp{\rm relatively\sp prime
\sp to\sp}k', \sp \Rightarrow \sp\{\ell a\}+\{\ell b\}<k',&\cr
k'/2<\{\ell\} <k',&\sp\ell\sp{\rm relatively\sp prime
\sp to\sp}k', \sp\Rightarrow\sp \{\ell a\}+\{\ell b\}>k'.&(4.1c)\cr}$$

\medskip\noindent{\bf Lemma 4 ($\rho$-coupling)}:\quad The
only solutions to eqs.(4.1$a,c$) are:

\item{(i)} for $k\equiv 2,4,7,8,10,11$ (mod 12):
$$(a,b)\in\{(1,1)\};\eqno(4.2a)$$

\item{(ii)} for $k\equiv 1,5$ (mod 12):
$$(a,b)\in \{(1,1),({k+1\over 2},{k+1\over 2})\};\eqno(4.2b)$$

\item{(iii)} for $k\equiv 0,3,6$ (mod 12):
$$(a,b)\in \{(1,1),(1,k+1),(k+1,1)\};\eqno(4.2c)$$

\item{(iv)} for $k\equiv 9$ (mod 12), $k\ne 21,57$:
$$(a,b)\in \{(1,1),(1,k+1),(2,{k+1\over 2}),(k+1,1),({k+1\over 2},2),
({k+1\over 2},{k+1\over 2})\}.\eqno(4.2d)$$

We will say $k$ is in class (i) if $k\equiv 2,4,7,8,10,11$ (mod 12), in class
(ii) if $k\equiv 1,5$ (mod 12), etc. The only $k$'s missing from this
list are $k=21,57$, each of which have 12 possibilities for $(a,b)$.
For $k=21$, these are precisely the 12 weights in (2.7$g$) lying in the
block containing $\chi^{21}_{1,1}$ --- namely (1,1), (5,5), (7,7), $\ldots$,
(7,10). For $k=57$ these are given in (5.11). The reason $k=21$
and $k=57$ are singled out here turns out to be the same (see Claim 1, and the
proof of Claim 3) as the reason $k=10$ and $k=28$ are singled out in the
$\rho$-coupling Lemma for $A_1$ (see Sec.6 and the Appendix). Indeed,
$21+3=2(10+2)$ and $57+3=2(28+2)$.

We will prove Lemma 4 later in the section. For now let us consider what
would happen if it were true.
For example, for $k\equiv 2,4,7,8,10,11$ (mod 12), or $k=1$,
${\cal R}^k=\{\rho\}$; and for $k\equiv 1,5$ (mod 12),
${\cal R}^k\subseteq\{\rho,{k+1\over 2}\rho\}$.
Then for half of the possible
levels, we will have reduced the completeness proof to the classification
of the permutation invariants, and considerable information about the
remaining levels will have been deduced. Lemma 4 turns out to be
sufficient to complete the proof of Thm.1 for all $k$ (this is done in Sec.5).

Incidently, taking $\ell=-1$ in (2.9) shows that, \eg, $N_{1,1;1,k+1}
=N_{1,1;k+1,1}$, for any invariant $N$ and level $k$. We will need this and
many other consequences of $(2.9)$ in Sec.5.

Lemma 4 can be thought of as related to the $A_1$ completeness proofs in
\CIZ{} and \ROB, and the $A_2$ $k'$ prime proof in \RTW, though it was
obtained independently. However it captures the big advantage the
approach developed in this paper has over those older approaches: through
it we impose positivity from the start; because of it we avoid explicit
construction of the commutant.

Before trying to understand the somewhat lengthy proof
of Lemma 4 given below, it may be wise for the reader to consult the
related, but considerably simpler, proof given at the end of Sec.5 of \COMM{}
for $\rho$-coupling for level 1 of $C_n$, odd $n$, or the proof for
$\rho$-coupling of $A_1$, all levels, given in the appendix of this paper.
We will find a strong relationship between the $A_1$ proof, and that of
$A_2$. In particular we will need the following result, proven in the
appendix:

\medskip\noindent{{\bf Claim 1}}: Let $K>a$ be positive integers, and
$a$ be odd. Suppose that for all $0<\ell<K$, $\ell$ relatively prime to $2K$,
we have
$\{\ell a\}_{2K}<K$, where $\{x\}_y$ is the unique number congruent (mod $y$)
to $x$ satisfying $0\le \{x\}_y<y$. Then

\item{(a)} for $K$ odd, $a=1$;

\item{(b)} for $K$ even and $K\ne 6,10,12,30$, $a=1$ or $K-1$;

\item{(c)} for $K=6$, $a=1,3,5$; for $K=10$, $a=1,3,7,9$; for $K=12$, $a=1,5,7
,11$; and for $K=30$, $a=1,11,19,29$.

\medskip
\noindent{{\bf Claim 2}}: For any $k$ and any $(a,b)\in P^k$, $(a,b)$ satisfies
the parity condition (4.1$c$) iff $\omega(a,b)$ does. Moreover,
 if $(a,b)$ satisfies the condition
$$a^2+ab+b^2\equiv 3 \sp({\rm mod}\sp k'),\eqno(4.3)$$
then so will $\omega(a,b)$, and if 3 divides $k'$, then $(a,b)$ will satisfy
the norm condition $(4.1a$) iff $\omega(a,b)$ will. \medskip

$\omega$ is the outer automorphism defined in (2.2$b$).  The proof of
Claim 2 is a straightforward calculation. For example, if $\{\ell a\}+\{\ell
b\}<k'$, then $\{\ell k'-\ell a-\ell b\}+\{\ell a\}=k'-\{\ell a\}-\{\ell b\}
+\{\ell a\}=k'-\{\ell b\}<k'$.

Because of Claim 2, we will restrict our attention for the remainder of this
section to any weight $(a,b)\in P^k$ satisfying the parity condition $(4.1c)$
and the norm condition (4.3) (and if $k'\equiv 0$ (mod 3) the stronger norm
condition $(4.1a)$). By Claim 2 this set of possible $(a,b)$ is invariant
under the 6 outer automorphisms. At the conclusion of our arguments we will
have a finite set of solutions $(a,b)$ to $(4.1c$) and (4.3); it suffices
then to find those weights among them which satisfy $(4.1a)$.\bigskip

\noindent{{\bf Proof of Lemma 4 when 4 divides $k'$}} \qquad We learn from the
norm condition ($4.3$) that two of $a$, $b$ and $k'-a-b$ will be odd and
one will be even; from Claim 2 we may assume for now that $a$ and $b$ are
odd. Let $0<\ell<k'/2$, $\ell$ relatively prime to $k'$. Then $\ell'=\ell
+k'/2$ will also be relatively prime to $k'$ but will lie in the range
$k'/2$ to  $k'$. Then ($4.1c$) tells us
$$\{\ell a\}+\{\ell b\}<k'<\{\ell' a\}+\{\ell' b\}.\eqno(4.4)$$
But $a$ is odd, so $\{\ell' a\}=\{k'/2+\ell a\}$ equals
$k'/2+\{\ell a\}$ if $\{ \ell a\}<k'/2$, or $-k'/2+\{\ell a\}$ if $\{\ell a
\}>k'/2$. A similar comment applies to $b$. From (4.4) we now immediately
get that both $\{\ell a\},\{ \ell b\}<k'/2$. Thus, putting $K=k'/2$ we
read off from Claim 1 that the only possibilities for $a$ and $b$ are
1 and $(k+1)/2$, unless $k'=12,20,24,60$. From these we can also compute the
possibilities for $k'-a-b$. Eq.(4.1$a$) now suffices to reduce this list of
possibilities to those given in Lemma 4. \qquad {\bf QED} to classes
(ii) and (iv) \bigskip

Thus it suffices now to consider $k\equiv 0,2,3$ (mod 4). As before let $(a,b)
\in P^k$ be any weight satisfying (4.$1c$) and (4.3) (and (4.$1a$) if 3
divides $k'$). First we
will prove two useful results.

\medskip\noindent{{\bf Claim 3}}: For $k\equiv 0,2,3$ (mod 4), if $a=b$ then
$a=b=1$.

\noindent{{\it Proof}} \quad  Clearly $a<k'/2$. First consider $k'$ odd.
Let $M>0$ be the smallest integer for which $2^M<k'/2<2^{M+1}$. Similarly,
let $N\ge 0$ be the smallest integer for which $2^Na<k'/2<2^{N+1}a$.
Assume for contradiction that $a>1$. Then $0\le N<M$. Take $\ell=2^{N+1}<k'/2$.
Then we get $\{\ell a\}+\{\ell a\}=2\{2^{N+1}a\}>k'$, contradicting
(4.1$c$).

For $k'$ even, (4.3) says $a$ must be odd. We can now directly apply
Claim 1(a) with $K=k'/2$, to again get $a=1$. \qquad QED to Claim 3\medskip

\noindent{{\bf Claim 4}}: The greatest common divisors of $a$ and $k'$,
of $b$ and $k'$, and of $k'-a-b$ and $k'$, equal either 1 or 2.

\noindent{{\it Proof}} \quad
Suppose a prime $p\ne 2$ divides both $a$ and $k'$. Then (4.1$a$) implies
$p\ne 3$, and (4.3) that $b^2\equiv 3$ (mod $p$) --- \ie 3 is a quadratic
residue of $p$, so $p\ge 11$.

Let $\ell_m=1+mk'/p$, $m=0,1,\ldots,p-1$. Except possibly for one value of
$m$, call it $m_0$, each $\ell_m$ will be relatively prime to $k'$. Assume
$k'\ne p$; if $k'=p$ the ranges given below for $m$ will be slightly
different but otherwise the same argument holds.
Therefore, for $m=0,\ldots,{p-1\over 2}$ (except possibly for $m=m_0$),
$(4.1c)$ says $\{b+mbk'/p\}<k'-a$, and for $m={p+1\over 2},\ldots,p-1$
(except possibly $m=m_0$), $\{b+mbk'/p\}>k'-a$. Because $p\ge 11$, it can
be shown that these two inequalities can only be satisfied if $bk'/p\equiv
\pm k'/p$ (mod $k'$), \ie $b\equiv \pm 1$ (mod $p$), in which case $b^2
\equiv 1\not\equiv 3$ (mod $p$).

Therefore, $p=2$ is the only prime that can divide both $a$ and $k'$. Since
$(4.3$) shows 4 cannot divide both, the only possibilities for the gcd
are 1 or 2. The same calculation applies to gcd($b,k'$) and, using Claim 2,
to gcd$(k'-a-b,k'$).
\qquad QED to Claim 4

\bigskip \noindent{{\bf Proof of Lemma 4 for $k'$ odd}}\qquad
{}From Claim 2 we may assume $1\le a,b\le k'/2$.
It suffices to show $a=b=1$.

First take $\ell=(k'-1)/2$;
it is relatively prime to $k'$ and less than $k'/2$.
If $a$ is even, $\{\ell a\}=k'-a/2$, and if $a$ is odd, $\{\ell a\}=k'/2
-a/2$. Hence $\{\ell a\}+\{\ell b\}=ik'+(k'-a-b)/2$, where $i=1/2,1,3/2$
depending on whether 0, 1 or both of $a,b$ are even. But $i\ge 1$ contradicts
(4.1$c$). Therefore both $a$ and $b$ must be odd.

Eq.(4.3) tells us $\{a^2\}+\{ab\}+\{b^2\}=3+mk'$, for some integer $m$.
Since by definition $0\le \{\cdots\}<k'$, we have $m=0,1,$ or 2. But
$m=2$ would imply $\{a^2\}+\{ab\}=3+2k'-\{b^2\}>k'$,
which contradicts (4.1$c$) with $\ell=a<k'/2$, by Claim 4.

Next suppose $m=1$, \ie
$$\{a^2\}+\{ab\}+\{b^2\}=k'+3.\eqno(4.5)$$
Choose $\ell=
(k'+a)/2$, $\ell'=(k'+b)/2$ --- again Claim 4 tells us these are relatively
prime to $k'$. Then $\ell a\equiv k'/2+a^2/2$ (mod $k'$),
so $\{\ell a\}=\{a^2\}/2+k'/2$ if $\{a^2\}$ is odd, and $\{a^2\}/2$ if
$\{a^2\}$ is even. Similarly, $\{\ell b\}=\{\ell'a\}=\{ab\}/2+k'/2$ or
$\{ab\}/2$, depending on whether $\{ab\}$ is odd or even, resp., and
$\{\ell'b\}=\{b^2\}/2+k'/2$ if $\{b^2\}$ is odd, and $\{b^2\}/2$ if $\{
b^2\}$ is even. But (4.5) tells us that $\{a^2\}+\{ab\}+\{b^2\}$ is even,
so either all three are even, or 2 are odd and 1 is even. If $\{a^2\}$ or
$\{ab\}$ are even, then using $\ell$ in (4.1$c$) gives $k'<\{a^2\}+\{ab\}$,
contradicting $a<k'/2$; otherwise using $\ell'$ contradicts $b<k'/2$.

Thus $m\ne 1$, so $m=0$ is forced. This gives us
$a^2\equiv ab\equiv b^2\equiv 1$ (mod $k')$; Claim 4 then
implies $a\equiv b$ (mod $k'$), which Claim 3 tells us forces $a=b=1$.
\qquad {\bf QED} to Lemma 4 for $k'$ odd \bigskip

\noindent{{\bf Proof for Lemma 4 for $k'\equiv 2$ (mod 4)}} \qquad This is the
final, and messiest, possibility; its proof uses tools resembling those
in the Appendix. From $(4.3)$ we get that both $a$ and $b$
cannot be even, so by Claim 2 we may assume $a,b$ are both odd.
Define $M$ by $2^M<k'/2<2^{M+1}$, so $k'/2^M<4$. Let us
begin with a useful fact.

\medskip\noindent{{\bf Claim 5}}: $a=1$ implies $b=1$.

\noindent{{\it Proof}} \quad The norm condition (4.3) becomes
$$b^2+b\equiv 2\sp ({\rm mod }\sp k').\eqno(4.6)$$
Take first $\ell=b$ in $(4.1c$); from (4.6) we get either $b=1$ or $b>k'/2$.
Suppose $b>k'/2$, and write $b=k'/2+b'$, so $b'$ is even and $0<b'<k'/2$.
Define $N$ so that $k'/2<2^Nb'<k'$. Then $0<N\le M$. Taking $\ell=k'/2+2^N$,
we get $k'<\{k'/2+2^N\}+\{(k'/2+2^N)b\}=k'/2+2^N+2^Nb'-k'/2=2^N+2^Nb'$.

Now take $\ell=(k'/2+2^N)b$. We get $k'>\{(k'/2+2^N)b\}+\{(k'/2+2^N)b^2\}=
2^Nb'-k'/2+\{k'/2+2^{N+1}-2^Nb'\}$, using (4.6). This forces $k'/2+2^{N+1}
-2^Nb'>0$. Hence $k'<2^N+2^Nb'<k'/2+2^N+2^{N+1}$, \ie $k'<3\cdot 2^{N+1}$,
so $b'<k'/2^N<6$, which tells us either $b'=2$ or $b'=4$.

It is easy to verify that $b=k'/2+2$ cannot satisfy (4.6), and $b=k'/2+4$
can only if $20\equiv 2$ (mod $k'$), \ie $k'=18$ or 6. These values can be
individually checked. \qquad QED to Claim 5 \medskip

Thus by Claims 3 and 5 it suffices to show there can be no solutions $(a,b)\in
P^k$ to
$(4.1c$) and (4.3) for $a,b$ odd, $1<a<b$. Write out the binary expansions
$a/k'=\sum_{i=1}^\infty a_i 2^{-i}$, $b/k'=\sum_{i=1}^\infty b_i 2^{-i}$,
where each $a_i,b_i\in \{0,1\}$.

Consider $\ell_i=k'/2+2^i$, $i=1,\ldots,M$. Then
$$k'<\{\ell_i a\}+\{\ell_i b\}=\{{k'\over 2}+2^ia\}+\{{k'\over 2}+2^ib\}=
\{2^ia\}+\{2^ib\}+\left\{\matrix{k'&{\rm if}\sp a_{i+1}=b_{i+1}=0\cr
0&{\rm if}\sp a_{i+1}+b_{i+1}=1\cr -k'&{\rm if}\sp a_{i+1}=b_{i+1}=1\cr}
\right. \eqno(4.7a)$$
But $\{\cdots\}<k'$, so $(4.7a)$ forbids $a_{i+1}=b_{i+1}=1$, for all
$i=1,2,\ldots,M$ (the relation $a+b<k'$ forbids it for $i=0$).

Define $I$ by $k'/2^I<b<k'/2^{I-1}$, \ie $b_i=0$ for $i<I$ and $b_I=1$.
Consider first the case $I>1$. Then ($4.7a$) tells us $k'<\{2^{I-1}a\}
+\{2^{I-1}b\}=2^{I-1}a+2^{I-1}b$, \ie $k'/2^{I-1}<a+b$. This strong
inequality now forces $a_i+b_i=1$ for $I\le i<M+1$, \ie
$$I>1\Rightarrow a+b={k'\over 2^{I-1}}+\epsilon,\sp {\rm where}\sp 0<
\epsilon<2.\eqno(4.7b)$$

The case $I=1$ is similar. Define $I'>1$ to be the smallest index (other
than $I=1$) with $a_{I'}=1$ or $b_{I'}=1$. Then the identical argument gives
$$I=1\Rightarrow a+b={k'\over 2}+{k'\over 2^{I'-1}}+\epsilon,\sp {\rm where}
\sp 0<\epsilon<2.\eqno(4.7c)$$
In both $(4.7b,c$), $\epsilon$
is fixed by the constraint that $a+b$ must be even. Thus we have essentially
removed one degree of freedom. First we will eliminate $I,I'=2,3$.

\medskip \noindent{\bf Claim 6}:\quad Either $I>3$, or $I=1$ and $I'>3$.
\medskip

\noindent{\it Proof}\quad Suppose first that $I=2$. Then $a+b=k'/2+1$, so
$\{ab\}=k'/2-2$, $\{a^2\}=a+2$, $\{b^2\}=b+2$. Therefore either ${a(a-1)\over
2}\equiv 1$ (mod $k'$) (if $a\equiv -1$ (mod 4)), or ${a(a-1)\over 2}+
{a\over 2}k'\equiv 1$ (mod $k'$) (if $a\equiv +1$ (mod 4)). Then $a\equiv
+1$ (mod 4) would violate $(4.1c$) with $\ell={a-1\over 2}+{k'\over 2}$, so
$a\equiv -1$ (mod 4). Similarly, we must have $b\equiv -1$ (mod 4), so
${k'\over 2}+1\equiv 2$ (mod 4), \ie $k'\equiv 2$ (mod 8). Now take $\ell=
{k'+2\over 4}$; we get $\{{-k'\over 2}+{a\over 2}\}+\{{-k'\over 4}+{b\over 2}
\}={3k'\over 2}+{k'+2\over 4}>k'$, contradicting $(4.1c$).

Now suppose $I=3$, \ie
$$a<{k'\over 8}<b<{k'\over 4},\quad a+b={k'\over 4}+\left\{ \matrix{{1\over
2}&{\rm if}\sp k'\equiv -2\sp({\rm mod}\sp 8)\cr {3\over 2}&{\rm if}\sp k'
\equiv +2\sp ({\rm mod}\sp 8)\cr}\right. . $$
Taking $\ell=k'/2-a-b$ eliminates $a\equiv b\equiv 3$ (mod 4). Perhaps the
easiest way to handle $I=3$ is to address the 4 possibilities for $k'$
(mod 16) individually. For $k'\equiv 2,6$ (mod 16) it turns out
$\{a^2\}+\{ab\}>k'$, violating $(4.1c$) with $\ell=
a$. For $k'\equiv 10$ (mod 16), $a\not\equiv b$ (mod 4) so take $\ell={k'\over
4}+{1\over 2}$.

The harder possibility is $k'\equiv 14$ (mod 16). Then $a+b={k'\over 4}+{1
\over 2}\equiv 0$ (mod 4), so $a\not\equiv b$ (mod 4). Here we have $({k'
\over 4})^2\equiv {7\over 8}k'$ (mod $k'$), so for $k'>14$
$\{ab\}={k'\over 8}-{11\over
4}$ and $\{b^2\}$ equals either ${5\over 8}k'+{b\over 2}+{11\over 4}$ or
${1\over 8}k'+{b\over 2}+{11\over 4}$. Taking $\ell={3\over 2}k'-4b$ gives
the contradiction $11+k'-2b-11>k'$. For the remaining case, $k'=14$, choosing
$\ell=a$ gives a contradiction.

Now suppose $I=1$. Then by $(4.7c$), $I'=2$ would violate $a+b<k'$. $I'=3$
can be handled similarly to $I=3$.  \qquad QED to Claim 6\medskip

Write $k'=2\cdot 3^L\cdot k''$, where $k''\equiv \pm 1$ (mod 6). Consider
first the case $L=0$. Define $J$ by ${k'\over 3\cdot 2^J}<b<{k'\over 3\cdot
2^{J-1}}$ if $I>1$, and if $I=1$ define $J$ to be the smallest number such that
either ${k'\over 3\cdot 2^J}<a<{k'\over 3\cdot 2^{J-1}}$ or
${k'\over 3\cdot 2^J}<b'<{k'\over 3\cdot 2^{J-1}}$.
Note that $J=I-1$ or $I-2$, and $J=I'-1$ or $I'-2$, in the 2 cases.
By Claim 6 we know $J>1$.
Putting $\ell'_i=k'/2-3\cdot 2^i$ into $(4.1c$) presents us with
a familiar calculation:
$$\eqalignno{I>1\Rightarrow\sp&a+b={k'\over 3\cdot 2^{J-1}}+\epsilon';&(4.8a)
\cr
I=1\Rightarrow\sp&a+b={k'\over 2}+{k'\over 3\cdot 2^{J-1}}+\epsilon';&(4.8b)\cr
}$$
where $0<\epsilon'<2$. Equating eqs.($4.7b$) with (4.8$a$) gives us $k'/(3
\cdot 2^{I-1})=\epsilon'-\epsilon$ if $J=I-1$, and $k'/(3\cdot 2^{I-1})=
\epsilon-\epsilon'$ if $J=I-2$. In either case we get $a<k'/2^I<3$, \ie
$a=1$. The $I=1$ calculation is identical.

Now consider $L>1$. We get from (4.3) that $a\equiv b$ (mod 3). $a\equiv -1$
(mod 3) is dealt with using $\ell=k'/6-2$, so $a$ must be $\equiv +1$ (mod
3). We will do $I>1$ (the proof for $I=1$ is similar).

An easy calculation from $a>1$ shows $2^{I-1}<k'/3$. Therefore both $\ell'
=k'/6+2^{I-2}$ and $\ell''=k'/6+2^{I-1}$ are less than $k'/2$, and $\ell'''
=k'/6-2^{I-3}$ is positive. $\ell'$ gives $a+b<2k'/(3\cdot 2^{I-2})$, while
$\ell''$ implies $2^{I-1}b+k'/6>k'$. Now $\ell'''$ yields $k'>k'/6-2^{I-3}a
+7k'/6-2^{I-3}b$, contradicting the $\ell'$ inequality.

Finally consider $L=1$. As before we will only give the proof for $I>1$ ---
it is similar for $I=1$. As before, we can force $a\equiv b\equiv +1$ (mod
3).

Define $J$ by $k'/2^J<3b<k'/2^{J-1}$. Then $J=I-1$ or $I-2$, so $J>1$ by
Claim 6. Assume for now that $J>2$; then $\ell'=k'/6+3\cdot 2^{J-2}$ and
$\ell''=k'/6+3\cdot 2^{J-1}$ give us $a+b<2k'/(9\cdot 2^{J-2})$ and
$b>5k'/(18\cdot 2^{J-1})$. The former tells us $J=I-2$, while the latter
demands $J=I-1$.

The remaining possibility, namely $J=2$ and $I=4$, is eliminated by taking
$\ell=\ell''=k'/6+6$. \qquad {\bf QED} to Lemma 4 for $k'\equiv 2$ (mod 4)

\vfill \eject

\centerline{{\bf 5. The remaining levels}}
\bigskip

In Sec.4 we concluded the proof that eqs.(2.7)
exhaust all physical invariants, for half the levels. In this section we will
conclude the $A_2$ classification problem for the remaining levels.

Until now the only properties of the partition functions we have exploited are
(P1), (P2) and (P3): the
invariants are {\it physical} invariants. However there are other conditions
known to be satisfied by the partition functions of all (unitary,
CPT-invariant)
conformal field theories. We will make these extra properties explicit
in the following two paragraphs; any physical invariant satisfying them
will be called a {\it strongly physical invariant}.

\MS{} tells us that to any level $k$ strongly physical invariant
$Z$ there are associated
two maximally extended chiral algebras, ${\cal A}$ and ${\bar {\cal A}}$,
which may or may not be isomorphic. These algebras are extensions of the
affine algebra ${\hat A_2}$; they both equal ${\hat A_2}$ iff $Z$ is a
permutation invariant. Let $ch_i$ and $\bar{ch}_j$ be their
characters.
These can be written as finite linear combinations
$$ch_i=\sum_{\la\in P^k} m_{i\la} \chi^k_{\la},\sp \sp \bar{ch}_j=
\sum_{\la\in P^k} {\bar m}_{j\la}\chi^k_{\la}\eqno(5.1a)$$
of characters $\chi^k_\la$ of ${\hat A}_2$, where the coefficients $m_{i\la},
{\bar m}_{j\la}$ are non-negative integers.
Let $ch_0$ and $\bar{ch}_0$ be the unique ones with $m_{0\rho},
{\bar m}_{0\rho}\ne 0$. ${\cal A}$ and ${\bar {\cal A}}$ must have an equal
number $n_c$ of characters. Then
$$Z=\sum_{i=0}^{n_c-1} ch_i \,\bar{ch}_{\pi i}^*,\eqno(5.1b)$$
for some permutation $\pi$ of the indices $\{0,\ldots,n_c-1\}$. In other words,
every strongly physical invariant is a sort of permutation invariant when
the chiral algebras are maximally extended.

Consider the matrices $S^e_{ij}$ and ${\bar S}^e_{ij}$ which describe the
behaviour of the extended characters $ch_i$ and $\bar{ch}_i$, respectively,
under the transformation $\tau\rightarrow-1/\tau$, as in (2.5$d$).
Then we know from \MS{} that $S^e$ and ${\bar S}^e$ are both unitary and
symmetric, and
$$S^e_{0j}\ge S^e_{00}>0, \qquad{\bar S}^e_{0j}\ge {\bar S}^e_{00}>0.
\eqno(5.2)$$

Now consider any level $k$ strongly physical invariant $Z$, given by
$(5.1b)$. One
immediate consequence of the above comments is that the function
$$Z'=\sum_{i=0}^{n_c-1} |ch_i|^2\eqno(5.3a)$$
also is a physical invariant. We will call an invariant of this form (\ie
diagonal in the extension) a {\it block-diagonal}. In a sense to be made clear
later, this observation will
allow us to simplify our arguments by permitting us to consider the existence
(or non-existence) solely of physical invariants of the form (5.3$a$).
Note that the coefficient matrix $N'$ of $Z'$ in (5.3$a$) is symmetric and
must satisfy
$$(N'_{\la\la'})^2\le N'_{\la\la}\,N'_{\la'\la'},\quad \forall \la,\la'\in
P^k. \eqno(5.3b)$$

Consider any $\la\in P^k$, and compute the sum
$$s(\la,{\cal A})\equi \sum_{\la'\in P^k}m_{0\la'}S^{(k)}_{\la'\la}=
\sum_{i=0}^{n_c-1}m_{i\la} S^e_{0i}.\eqno(5.4)$$
The second equality here follows because $S^e$ is symmetric. But $m_{i\la}
\ge 0$ and by (5.2) $S^e_{0i}>0$, so the RHS of (5.4) is non-negative and
will be zero only if all $m_{i\la}=0$. This gives us a simple but powerful
test for the character $ch_0$. In particular, we can read off from
Lemma 4 the possibilities for $ch_0$; most of these will have
$s(\la)<0$ or $s(\la)$ non-real for some $\la\in P^k$ and so can be
dismissed.

Hence our argument will depend crucially on Lemma 4, so will be broken
down into 4 cases: class (ii); class (iii); class (iv); and the exceptional
value $k=57$.

\medskip {\noindent \underbar{Class (ii)}}

Consider first class (ii), \ie all $k\equiv 1,5$ (mod 12). We may consider
$k>1$, because by Lemmas 4 and 1, any physical invariant at $k=1$ is a
permutation invariant and thus is enumerated in Thm.2. Suppose there
exists a level $k$ strongly physical invariant which is not a permutation
invariant. We would like to show that, except for $k=5$, this cannot
happen. Write $\rho'=({k+1\over 2},{k+1\over 2})$. Lemma 4 tells us that
$N_{\rho,\la}=N_{\la,\rho}=0$ except for $N_{\rho,\rho}=1,$ and $N_{\rho,
\rho'},N_{\rho',\rho}$. Write $ch_0=\chi^k_\rho+a\chi^k_{\rho'}$, $\bar{ch}_0
=\chi^k_\rho+b\chi^k_{\rho'}$, for non-negative integers $a=N_{\rho',\rho}$, $
b=N_{\rho,\rho'}$. At least one of $a,b$ must be non-zero (otherwise by
Lemma 1 $N$ would be a permutation invariant) --- without loss of generality
say $a>0$. Then the corresponding block-diagonal (5.3$a$) will also be a
physical non-permutation invariant of level $k$. Let us then assume our
invariant is in block-diagonal form. If we can show there is no
block-diagonal invariant corresponding to these $ch_i$,
we will have succeeded in showing no strongly physical invariant can exist
at these levels unless it is a permutation invariant, and we will have
completed the proof of Thm.1($b)$ for these levels.

{}From (5.3$b$) we get $N_{\rho',\rho'}\ge a^2$, where the inequality
will hold iff $m_{i\rho'}\ne 0$ for some $i>0$. However, taking
$\la_L=\la_R=\rho$ and $\ell={k+1\over 2}$ in eq.(2.9) tells us
$1=N_{\rho,\rho}=N_{\rho',\rho'}$, so $a=1$ and $m_{i\rho'}=0$ for all $i>0$.

Taking $\la=(1,2)$, note that eq.(2.5$f$) gives us
$$s(\la)=S^{(k)}_{\rho\la}+S^{(k)}_{\rho'\la}={4\over k'\sqrt{3}}\{
\sin[2\pi /k']-\sin[6\pi /k']\}.$$
$s(\la$) equals 0 for $k=5$, but is negative
for all larger $k\equiv 1$ (mod 4).  By the discussion after (5.4), this means
no strongly physical invariant (except possibly for $k=5$) can have
$ch_0=\chi^k_\rho+\chi^k_{\rho'}$,
which concludes the proof of Thm.1(b) for class (ii).

\medskip{\noindent \underbar{Class (iii)}}

Class (iii), \ie $k\equiv 0,3,6$ (mod 12), is more difficult. As in class
(ii), it suffices to consider block-diagonal invariants. From Lemma 4 we
read off $ch_0=\chi^k_\rho+a\chi^k_{\rho'}+b\chi^k_{\rho''}$, where now we
take $\rho'=(k+1,1)$ and $\rho''=(1,k+1)$, and where at least one of
$a,b$ is positive. Taking $\ell=-1$ and $(\la;\la')=(\rho;\rho')$ in
(2.9) tells us $a=b\ge 1$. We would first like to show $a=1$.

$\omega(\rho)=\rho'$ and $\omega^2(\rho)=\rho''$, where $\omega$ is defined
in (2.2). Put $\la'=(m,n)$; then for any $\la$ we get
from (2.5$f$)
$$S^{(k)}_{\omega(\la),\la'}=\exp[2\pi i(m-n)/3]\,S^{(k)}_{\la\la'},\quad
S^{(k)}_{\omega^2(\la),\la'}=\exp[2\pi i(-m+n)/3]\,S^{(k)}_{\la\la'}.
\eqno(5.5a)$$
Substituting in $\la'=(1,2)$, we find that
$$S^{(k)}_{\rho,(1,2)}+aS^{(k)}_{\rho',(1,2)}+aS^{(k)}_{\rho'',(1,2)}=
(1-a)\,S^{(k)}_{\rho,(1,2)},\eqno(5.5b)$$
where $S^{(k)}_{\rho,(1,2)}>0$. For $a>1$ $s(\la)$
will be negative, so we must have $a=1$.

Similar calculations show that $m_{i,(m,n)}$ will be zero for all $i=0,1,
\ldots,n_c-1$ iff $m\not\equiv n$ (mod 3), and that $m_{i,\rho'}=m_{i,\rho''}
=0$ for all $i=1,\ldots,n_c-1$.

We can partition the indices $\{0,\ldots,n_c-1\}$ into disjoint sets $I_\ell$,
where $i,j$ lie in the same set $I_\ell$ iff there exists a $\la\in P^k$
such that
$m_{i\la},m_{j\la}>0$. For example we have just shown that one set, call it
$I_0$, equals $\{0\}$. Rewrite (5.3$a$) as
$$Z=\sum_\ell \sum_{i\in I_\ell} |ch_i|^2. \eqno(5.6a)$$
This is equivalent to writing $N$ as a direct sum of indecomposable
matrices $N_\ell$ as in (2.10$c$), where
$$N_0=\left(\matrix{1&1&1\cr 1&1&1\cr 1&1&1\cr}\right)\equi M_{3,1}
\eqno(5.6b)$$
is the block `containing' $\chi^k_\rho$, $\chi^k_{\rho'}$ and
$\chi^k_{\rho''}$. What are the possibilities for the other $N_\ell$?
The following result is the heart of the class (iii) proof.

\medskip\noindent{\bf Claim}:\quad $N$ can be written as a direct sum of
matrices $N_\ell$, where either
$$N_\ell=(0)\equi M_{1,0},\qquad N_\ell=(3)\equi M_{1,3}\qquad{\rm or}\qquad
N_{\ell}=M_{3,1}.\eqno(5.6c)$$
In other words, each $ch_i$ either equals $\chi^k_\la$ for some $\la$ (in
which case there also are $i',i''\ne i$ for which $ch_{i'}=ch_{i''}=ch_i$),
or $ch_i=\chi^k_{\la_1}+\chi^k_{\la_2}+\chi^k_{\la_3}$ for some distinct
weights $\la_1,\la_2, \la_3$.
Moreover, $ch_i=\chi^k_\la$ can only happen for $\la=(k'/3,
k'/3)$, and $ch_i=\chi^k_{\la_1}+\chi^k_{\la_2}+\chi^k_{\la_3}$ can only
happen
for $\la_2=\omega(\la_1)$ and $\la_3=\omega^2(\la_1)$, up to a possible
reordering of the $\la_i$.

\medskip
\noindent{\it Proof}\quad From (5.6$b$) and (2.10$a$) we find
$r(N_0)=3$.
It is an easy combinatorial exercise to find all possibilities for $N_\ell$
with $r(N_\ell)=3$:  $N_\ell$ equals either
$$M_{1,3},\sp M_{3,1},\sp \left(\matrix{2&1\cr 1&2\cr}\right)\equi
M', \sp {\rm or}
\sp \left(\matrix{1&1&0\cr 1&2&1\cr 0&1&1}\right)\equi M'' .\eqno(5.6d)$$
The main things to keep in mind when showing (5.6$d$) is complete are
eqs.(2.10), and the fact that $N_\ell$ is the coefficient matrix for
$\sum_{i\in I_\ell}|ch_i|^2$. For example, if any entry of
$N_\ell$ is at least
3, then by (5.3$b$) a diagonal entry of $N_\ell$ is at least 3, so by
(2.10$b$) $N_\ell$ must equal $(3)=M_{1,3}$.

We wish to show that no $N_\ell$ can equal either $M'$ or $M''$. Since
$M_{1,3}^2=3M_{1,3}$ and $M_{3,1}^2=3M_{3,1}$, Lemma 3 would then conclude the
proof of the first statement of the claim.

First let us make some general remarks. Because of $(5.6d$), we know all of
the possible extended characters
$ch$ look like $ch=\sum_{j=1}^h \chi_{\la_j}^k$ for $h=1$, 2 or 3, where each
$\la_j=(m_j,n_j)\in P^k$ is distinct. Then we know
$$\sum_{j=1}^h S^{(k)}_{m_jn_j;m'n'}=0\eqno(5.7a)$$
must hold for each $(m',n')\in P^k$ with $m'\not\equiv n'$ (mod 3).
In fact, because the {\it triality} $m'-n'$ (mod 3) is preserved both by
Weyl reflections and adding vectors in $k'A_2$, we may drop the assumption
in (5.7$a$) that $(m',n')\in P^k$ and demand only that $m'\not\equiv n'$
(mod 3) (for $(m',n')\not\in P^k$ define $S^{(k)}_{mn,m'n'}$ by $(2.5f$)).

Writing $(m',n')=(\ell+\ell',\ell)$, $x_\ell=\exp[-2\pi i \ell/k']$ and
$y_{\ell'}=\exp[-2\pi i \ell'/3k']$, $(5.7a)$ becomes
$$\eqalignno{0=y_{\ell'}^{2k'}x_\ell^{k'}
\sum_{j=1}^h \{& y_{\ell'}^{2m_j+n_j}x_\ell^{m_j+n_j}+
y_{\ell'}^{-m_j+n_j}x_\ell^{-m_j}+y_{\ell'}^{-m_j-2n_j}x_\ell^{-n_j}&\cr
&-y_{\ell'}^{-m_j-2n_j}x_\ell^{-m_j-n_j}-y_{\ell'}^{2m_j+n_j}x^{m_j}_\ell
-y_{\ell'}^{-m_j+n_j}x^{n_j}_\ell\},&(5.7b)\cr}$$
where the extra irrelevant factor in front is added for future convenience
(to make the exponents all positive).
This must hold for all $\ell,\ell'\in \Z$, with $\ell'\equiv 0$ (mod 3).
Consider the polynomial $p(x,y)$ obtained by replacing $x_\ell$ and
$y_{\ell'}$ with the variables $x$ and
$y$, respectively, in the RHS of $(5.7b$). Divide $x^{k'}-1$ into $(y^{k'}-1)
p(x,y)$; the remainder is
$$\eqalignno{\sum_{j=1}^h&\{y^{3k'-m_j+n_j}x^{k'-m_j}-y^{2k'-m_j+n_j}
x^{k'-m_j}+y^{3k'-m_j-2n_j}x^{k'-n_j}&\cr &-y^{2k'-m_j-2n_j}x^{k'-n_j}-
y^{3k'-m_j-2n_j}x^{k'-m_j-n_j}+y^{2k'-m_j-2n_j}x^{k'-m_j-n_j}&\cr
&+y^{3k'+2m_j+n_j}x^{m_j+n_j}-y^{2k'+2m_j+n_j}x^{m_j+n_j}-y^{3k'+2m_j+n_j}
x^{m_j}&\cr &+y^{2k'+2m_j+n_j}x^{m_j}-y^{3k'-m_j+n_j}x^{n_j}+y^{2k'-m_j+n_j}
x^{n_j}\}.&(5.7c)\cr}$$
Then $(5.7b$) is equivalent to the statement that $y^{3k'}-1$ must divide
$(5.7c$).

Let us look at one of these terms, say $y^{2k'+2m_j+n_j}x^{m_j}$. Because
$y^{3k'}-1$ must divide the polynomial in (5.7$c$), there are only
two possibilities: either that one of the $6\cdot h$ terms in ($5.7c$) with
a coefficient of $-1$ will cancel $y^{2k'+2m_j+n_j}x^{m_j}$; or that one of
the
other $6\cdot h-1$ terms in $(5.7c$) with a coefficient of $+1$ will equal
$y^{2m_j+n_j-k'}x^{m_j}$. Let us first show that the second possibility
cannot be realized.

Suppose \eg that $y^{2k'-m_i-2n_i}x^{k'-m_i-n_i}=y^{2m_j+n_j-k'}x^{m_j}$, for
some $1\le i\le h$. That means $2k'-m_i-2n_i=2m_j+n_j-k'$ and
$k'-m_i-n_i=m_j$,
\ie $2k'-n_i=m_j+n_j$. But this contradicts $n_i<k'$ and $m_j+n_j<k'$. The
other possibilities all fail for similar reasons.

Thus the first possibility must be realized. Suppose \eg that
$y^{2k'-m_i+n_i}x^{k'-m_i}=y^{2k'+2m_j+n_j}x^{m_j}$, \ie $2k'-m_i+n_i=2k'+
2m_j+n_j$ and
$k'-m_i=m_j$. This gives us $-k'+n_i=m_j+n_j$, which is likewise impossible.
Indeed, the only positive terms in $(5.7c$) which can cancel
$y^{2k'+2m_j+n_j}x^{m_j}$ are $y^{3k'-m_i+n_i}x^{n_i}$ for any $i$, which
give
us the equations $n_i=m_j$ and $m_i=k'-m_j-n_j$. In other words, for each
$j$ there must be an $i$ such that $\omega(m_j,n_j)=(m_i,n_i)$.

Suppose $h=1$, \ie $ch=\chi^k_{m,n}$. Then $\omega(m,n)=(m,n)$, which can
only happen for $(m,n)=(k'/3,k'/3)$.

Suppose $h=2$. Then either $\omega(m_1,n_1)=(m_1,n_1)$ and $\omega(m_2,n_2)
=(m_2,n_2)$, or $\omega(m_1,n_1)=(m_2,n_2)$ and $\omega(m_2,n_2)=(m_1,n_1)$.
In either case, the only way this can happen is if $(m_1,n_1)=(m_2,n_2)=
(k'/3,k'/3)$, contradicting $\la_1\ne \la_2$. Therefore $h$ cannot equal 2.
Hence $N_\ell=M'$ and $N_\ell=M''$ are both impossible, because
both require an extended character with $h=2$.

Finally, suppose $h=3$. It is easy to verify that the only possibility
here is $\omega(m_1,n_1)=(m_2,n_2)$ and $\omega(m_2,n_2)=(m_3,n_3)$,
relabelling the indices if necessary.
\qquad QED to claim\medskip

{}From the claim, and the earlier observation that $m_{i,(m,n)}\neq 0$ for some
$i$ iff $m\equiv n$ (mod 3), we know already what our block-diagonal
invariant must look like: it is ${\cal D}_k$. Note that this extended algebra,
which we will call $A^e_{2,k}$,
has far fewer characters than $A_{2,k}$ does, so there are no `hybrid'
invariants with \eg the chiral algebras ${\cal A}=A^e_{2,k}$ and
${\bar {\cal A}}=A_{2,k}$: the only possibilities for a strongly physical
invariant are ${\cal A}={\bar {\cal A}}=A_{2,k}$ and ${\cal A}=
{\bar {\cal A}}=A^e_{2,k}$. The first possibility corresponds to permutation
invariants. Our task here is to enumerate all physical invariants
corresponding to the second possibility, in other words to find all
permutations $\pi$ of the extended characters which obey
$T^e_{\pi i,\pi j}=T^e_{ij}$ and
$$S^e_{\pi i,\pi j}=S^e_{ij}.\eqno(5.8a)$$
Also, $N^e_{\pi h,\pi i,\pi j}=N^e_{hij}$ where
$$N^e_{hij}\equi \sum_{g=0}^{n_c-1} {S^e_{hg} S^e_{ig} S^e_{jg} \over
S^e_{0g}}.\eqno(5.8b)$$

Write $ch_{m,n}$ for $\chi^k_{m,n}+\chi^k_{\omega(m,n)}+\chi^k_{\omega^2(m,n)}$
and $ch_{(i)}$, $i=1,2,3$, for the extended characters equal to
$\chi^k_{k'/3,k'/3}$. To avoid redundancy, we may restrict $(m,n)$ here to
lie in the set $P^{e}=\{(m,n)\in P^k\mid m<k'-m-n$ and $n\le k'-m-n\}$.

Using $(5.5a)$ we can easily find most of the entries of $S^e_{ij}$:
$$\eqalignno{S^e_{mn,m'n'}=&3S^{(k)}_{mn,m'n'},&(5.9a)\cr
S^e_{mn;(i)}=&S^{(k)}_{mn;k'/3,k'/3} \quad i=1,2,3,&(5.9b)\cr}$$
using obvious notation. The 9 remaining entries, $S^e_{(i)(j)}=
S^e_{(j)(i)}$ for $i,j=1,2,3$, satisfy several relations: \eg
$$|S^e_{(i)(1)}|^2+|S^e_{(i)(2)}|^2+|S^e_{(i)(3)}|^2={2\over 3}+{|\alpha|^2
\over 3},\eqno(5.9c)$$
where $\alpha=S^{(k)}_{k'/3,k'/3;k'/3,k'/3}={6\over \sqrt{3}k'}
\sin(2\pi k'/9)$.
Eq.(5.9$c)$ follows from (5.9$b$) and the unitarity of $S^e$ and $S^{(k)}$.

It will not be necessary for us to explicitly compute the $S^e_{(i)(j)}$. It
suffices to note from $(5.9c$) that for each $i$, there is a $j$ for which
$|S^e_{(i)(j)}|\ge \sqrt{2}/3$. On the other hand, by $(5.9a,b$)
the other entries of $S^e$
are proportional to $1/k'$, and so will usually be much smaller.

Suppose $\pi$ takes some $ch_{(i)}$ to some $ch_{m,n}$. A quick calculation
shows $|S^e_{mn,m'n'}|\ge \sqrt{2}/3$ can only happen for $k\le 19$, \ie $k=
18,15,12,\ldots,3$. An explicit computer calculation shows $|S^{(18)}|
<\sqrt{2}/9$, which eliminates $k=18$. Similarly $|S^e_{mn,(j)}|\ge
\sqrt{2}/3$ can only happen for $k\le 3$. Therefore, by $(5.8a$) for
$k\ge 18$
we must have $\pi$ taking each $ch_{m,n}$ to some $ch_{m',n'}$. Let us
restrict
ourselves to these $k$. How $\pi$ permutes the $ch_{(i)}$ is irrelevant to
us, since those 3 characters are equal.

To find all such $\pi$ for $k'>18$ reduces to arguments familiar from
earlier parts of this paper, so we will only give a 3-step sketch of the
proof.

Let ${\cal S}^{e}_a=\{(m,n)\in P^{e}\mid m=a$ or $n=a\}$. Suppose we know
$\pi(a,b)\in {\cal S}^{e}_a\cup {\cal S}^{e}_b$. Then using $(5.9a$) the
argument surrounding eqs.(3.7) applies here and tells us that $\pi(a,b)=
(a,b)$ or $(b,a)$.

Because of $(5.9a,b)$ and $(5.5a$) we can see that
$$\eqalignno{N^e_{mn,m'n',m''n''}&=\sum_{i=0}^2
N^{(k)}_{\omega^i(mn),m'n',m''n''},\quad {\rm so}&(5.10a)\cr
N^e_{mn,mn,mn}&={\rm min}\{m,n\}+\left\{\matrix{m+2n-{2k'\over 3}&{\rm if}\sp
{k'\over 3}-{m\over 2}<n\le {k'\over 3}\cr {k'\over 3}+m-n&{\rm if}\sp
{k'\over 3}\le n<{k'\over 3}+m\cr 0&{\rm otherwise}\cr}\right. &\cr&+
\left\{\matrix{n+2m-{2k'\over 3}&{\rm if}\sp
{k'\over 3}-{n\over 2}<m\le {k'\over 3}\cr {k'\over 3}+n-m&{\rm if}\sp
{k'\over 3}\le m<{k'\over 3}+n\cr 0&{\rm otherwise}\cr}\right. .&(5.10b)\cr}$$

\noindent{\it Step 1} First prove that $\pi(1,a)=(1,a)$ or $(a,1)$ for all
$(1,a)\in P^e$.

\noindent{It} suffices to show $\pi(1,a)\in {\cal S}^e_1$. This follows
immediately from $(5.10b$), except for $a=k'/3$ when $k'/3\equiv 1$ (mod 3).
If $\pi(1,k'/3)\not\in {\cal S}^e_1$, then $\pi(1,a)=(2,m)$ or ($m,2)$ for
some $m$. Now, we can show $\pi(3,3)=(3,3)$ (otherwise by $(5.10b)$
it must equal
$(2,k'/3+1$) or $(k'/3+1,2)$, which violates $S^e_{11,33}=S^e_{11,\pi(33)}$).
Then $S^e_{33;1,k'/3}=S^e_{33;2m}$ implies $m=k'/3-2$, which violates
$S^e_{11;1,k'/3}=S^e_{11;\pi(1,k'/3)}$.

\noindent{\it Step 2} Next show that $\pi(1,4)=(1,4)$ implies $\pi(1,a)=(1,a)$
for all $(1,a)\in P^e$.

\noindent{The} proof is similar to that used in proving $(3.10a$). In
particular, $(3.9a$) becomes
$$c_k(6+9a)+c_k(9+3a)+c_k(3-6a)=c_k(9+6a)+c_k(6-3a)+c_k(3+9a),\eqno(5.10c)$$
which has the solutions $a= 0,1,k'-{1\over 2}$ in the range $0\le a<k'$.

\noindent{\it Step 3} Finally, show that $\pi(1,4)=(1,4)$ implies
$\pi(a,b)=(a,b)$ for all $(a,b)\in P^e$.

\noindent{This} argument resembles the one surrounding $(5.7c$). Step 2 can
be used to show that $x^{k'}-1$ must divide the polynomial
$$\eqalignno{[e_k(3a+3b)-e_k(3b)]x^{a+2b}&+[e_k(3b')-e_k(3a'+3b')]x^{a'+2b'}
&\cr+[e_k(-3a)-e_k(-3a-3b)]x^{-2a-b}&
+[e_k(-3a'-3b')-e_k(-3a')]x^{-2a'-b'}&\cr +[e_k(-3b)-e_k(3a)]x^{a-b}&
+[e_k(3a')-e_k(-3b')]x^{a'-b'},&(5.10d)\cr}$$
where $(a',b')=\pi(a,b)$. This can only happen for $(a',b')=(a,b)$.

The conclusion is that there are only two possible permutations $\pi$, except
possibly for $k\le 15$. These give rise to the physical invariants ${\cal D}_k$
and ${\cal D}_k^c$. (The only levels this
argument breaks down at are $k=3,6,9,12,15$.)

\medskip {\noindent \underbar{Class (iv)}}

{}From Lemma 4, and using the previous arguments, the only possibilities
for $ch_0$ are $\chi^k_\rho$; $\chi^k_{\rho}+a\chi^k_{k+1,1}+a\chi^k_{1,k+1}$;
$\chi^k_{\rho}+\chi^k_{(k+1)/2,(k+1)/2}$; and $\chi^k_{\rho}+a\chi^k_{k+1,1}
+a\chi^k_{1,k+1}+\chi^k_{(k+1)/2,(k+1)/2}+a\chi^k_{(k+1)/2,2}+a
\chi^k_{2,(k+1)/2}$, where $a\ge 1$. The first corresponds to a permutation
invariant, and is classified in Thm.2: the permutation invariants are
${\cal A}_k$ and ${\cal A}^c_k$. The second is classified (for $k\ne 9$)
by the class (iii)
argument given above: the physical invariants are ${\cal D}_k$
and ${\cal D}^c_k$ (it turns out that the exceptionals ${\cal E}^{(2)}_9$ and
${\cal E}^{(2)c}_9$ also correspond to this $ch_0$). The third
possibility is dealt with using the argument of class (ii) given earlier,
and is realized by no physical invariants.

The fourth and final possibility can be dealt with using the eq.(5.4)
argument. In particular, the vector $\la=(1,3)$ in (5.4)
implies $a=1$. Then the choice $\la=(1,4)$ gives a contradiction,
except for level $k=9$ (where there is an exceptional, ${\cal E}^{(1)}_9$,
which realizes this $ch_0$ possibility).

\medskip {\noindent \underbar{Level 57}}

Lemma 4 for $k=57$ is: if $N_{\rho\la}\neq 0$ for a positive invariant of
level 57, then
$$\eqalignno{\la\in\{&(1,1),(1,58),(2,29),(11,11),(11,38),(19,19),(19,22),
(29,29),&\cr &(22,19),(29,2),(38,11),(58,1)\}.&(5.11)\cr}$$
{}From here we can read off the possibilities for $ch_0$: there are 16 of them,
half of which involve a parameter $a\ge 1$. Of these, four were considered
in the class (iv) argument given above. The remaining 12 all
succumb to similar arguments: the weight (1,3) used in (5.4)
forces $a=1$ in the 6 remaining possibilities involving the parameter $a$;
the weight (2,5) eliminates the 6 not involving $a$, and eliminates $a=1$
in the other 6. The conclusion is that
the only $k=57$ strongly physical invariants are the permutation invariants
${\cal A}_{57}$, ${\cal
A}_{57}^{c}$, ${\cal D}_{57}$ and ${\cal D}^c_{57}$.

\bigskip \bigskip \centerline{{\bf 6. Extensions to other algebras}}
\bigskip

The main motivation for pursuing a classification proof for $A_2$ is
the hope that the methods developed there would also be of use for other
algebras. And indeed that should be the case. The main issues are the
simplicity of the form the $\rho$-coupling Lemma for those algebras takes,
and also how well we can manage finding all permutation invariants.
For some examples we will now write down the
$\rho$-couplings for $A_1$, and a little later on that of $A_1\oplus A_1$ for
relatively prime levels $k_1+2$, $k_2+2$, as well as
conjectures for $G_2$ and $C_2$
which we have verified on a computer for the first hundred
levels (the $\rho$-shifted weights are identified with their Dynkin labels):

\medskip\noindent{\bf $\rho$-coupling for $A_1$}:\quad (a) For $k\equiv 1,2,3$
(mod 4), $k\ne 10$, the
only possible weight $\lambda$ which can couple to $\rho=1$ in a positive
invariant is $\la=1$;

(b) for $k\equiv 0$ (mod 4), $k\ne 28$, the only possibilities are
$\lambda=1$ and $k+1$;

(c) for $k=10$, $\la=1,7$; and for $k=28$, $\la=1,11,19,29$.

\medskip\noindent{\bf $\rho$-coupling for $G_2$}:\quad (a*) For $k\ne 3,4$,
the only possible weight $\la$ which can couple to $\rho=(1,1)$ in a positive
invariant is $\la=(1,1)$;

(b) for $k=3$, the only possibilities are $\la=(1,1),(2,2)$; and for $k=4$,
$\la=(1,1),(4,1)$.

\medskip\noindent{\bf $\rho$-coupling for $C_2$}:\quad (a*) For $k$ odd,
$k\ne 3,7$, the only possible weight $\la$ which can couple to $\rho=(1,1)$
in a positive invariant is $\la=(1,1)$;

(b*) for $k$ even, $k\ne 12$, the only possibilities are $\la=(1,1),(1,k+1)$;

(c) for $k=3$, $\la=$(1,1), (3,2); for $k=7$, $\la=$(1,1),
(3,3), (1,6), (7,2); and for $k=12$, $\la=$(1,1),
(3,4), (7,1), (5,5), (9,2), (3,8), (9,4), (7,7), (1,13).\medskip

The results for $A_1$ are proven in the appendix.
As yet unproven results are marked with an `$*$' (but because they hold for
$k<100$, they likely hold for all $k$). This shows that
in many ways $A_2$ is the least tractible of the rank 2 algebras.
According to these findings, for $G_2$
all physical invariants (except at levels 3 and 4) must be permutation
invariants, and similarly for $C_2$ at odd levels (except 3 and 7).
Note that for each of these algebras, at any level with
irregular $\rho$-coupling behaviour (\eg $k=10$ and 28 for $A_1$)
exceptional invariants can always be found. The only known
exception to this rule is $A_2$ at $k=57$.

The $\rho$-coupling possibilities for $A_1\oplus A_1$ are more complicated,
because
there are now two independent levels $k_1$ and $k_2$. But it is easy to find
these when \eg $k_1+2$ and $k_2+2$ are relatively prime,
using an argument very similar to those used in the Appendix (see \GR{}
for a proof). The result is:

\medskip\noindent{\bf $\rho$-coupling for $A_1\oplus A_1$}\quad When
$k_1+2$ and
$k_2+2$ are relatively prime, the only $\la=(m,n)$ which can couple to
$\rho=(1,1)$ in some positive invariant are:

\item{(i)} for $k_1,k_2$ odd, $k_1\equiv k_2$ (mod 4), then $\la=(1,1)$;

\item{(ii)} for $k_1,k_2$ odd, $k_1\not\equiv k_2$ (mod 4), then
$\la=(1,1),(k_1+1,k_2+1)$;

\item{(iii)} for $k_1\equiv 0$ (mod 4), $k_1\ne 28$, then
$\la=(1,1),(k_1+1,1)$;

\item{(iv)} for $k_1\equiv 2$ (mod 4), $k_1\ne 10$, then $\la=(1,1)$;

\item{(v)} for $k_1=10$, $\la=(1,1),(7,1)$; and for $k_1=28$, $\la=(1,1),
(11,1),(19,1),(29,1)$.\medskip

Since we also found in \GR{} all permutation invariants for $A_1\oplus A_1$,
for all
levels $k_1,k_2$, it is now an easy task, using the techniques of Sec.5,
to complete the $A_1\oplus A_1$ classification when $k_1+2$, $k_2+2$ are
relatively prime. These observations have also been generalized in \GR{}
to all $A_1\oplus \cdots \oplus A_1$, when $k_1+2,\ldots,k_L+2$ are all
relatively prime.

These findings suggest that Lemma 2 continues to be useful for algebras
other than just $A_2$. Our proof in Sec.3 to find all permutation invariants
of $A_2$ made use of explicit formulas for the $A_2$ fusion
rules \BMW, and those do not exist at present for the other algebras
(except for $A_1$ \GW{} and hence all sums of $A_1$ and $A_2$).
Our hope is that this will not constitute a
serious stumbling block for future applications of these ideas. The only
fusion rules
which our proof crucially needed were $N_{\la\la\la}$, which may be
simple enough to calculate explicitly. Moreover, since all the information
obtainable from fusion rules is also encoded in the modular $S$-matrix \VER,
though in not so accessible a form, it is possible that
alternate proofs of Thm.2 can be found which do not require explicit
knowledge of any fusion rules of $A_2$.

Perhaps a more serious problem facing generalizations of these techniques
to higher ranks is the dependence of many steps on explicit knowledge
of the modular $S$-matrix. The Weyl group of the algebra increases
fantastically as the rank, so so will the complexity of the
explicit formula for the modular $S$-matrix.

It was proven in \VT{} that there is an exact rank-level duality between
$C_n$ level $k$ and $C_k$ level $n$; in particular there is a one-to-one
correspondence between the physical invariants of one and those of the
other. Thus finding all the physical invariants of $C_2$ would mean we have
also found all the level 2 invariants of $C_n$. There also is an approximate
rank-level duality between $A_n$ level $k$ and $A_{k-1}$ level $n+1$ \WAL.
This suggests that the situation for levels 2 and 3 of $A_n$
should be approximately as accessible as that for arbitrary levels of
$A_1$ and $A_2$.  This will be another direction for
our future research.

\bigskip\bigskip \centerline{{\bf 7. Comments}} \bigskip

In this paper we first find all {\it permutation invariants} of $A_2$, for
each level $k$. We then prove that for $k\equiv 2,4,7,8,10,11$ (mod 12),
the only
level $k$ {\it physical invariants} of $A_2$ are permutation invariants.
Together, these two statements allow us to write down all physical invariants
for $A_2$ of those levels: ${\cal A}_k,$ ${\cal A}^c_k$, ${\cal D}_k$ and
${\cal D}^c_k$ (see eqs.$(2.7a,b,h)$).

To handle the remaining levels, we make use of additional results known to
be satisfied by the partition functions \MS. These
allow us to find all strongly physical invariants for $k\equiv 0,1,3,5,6,9$
(mod 12), except for 7 levels which have
been completely treated by the computer program of \HK. Thus
the classification problem for $A_2$ modular invariant partition functions
has now been completed.

Two questions suggest themselves: (i) At present our only
proof for levels $k\equiv 0,1,3,5,6,9$ (mod 12) requires results from \MS;
although these must hold for the partition function of any physically
reasonable conformal field theory, they
do not necessarily hold for invariants satisfying only the three
conditions (P1), (P2) and (P3). It would be desirable to reduce as much as
possible the required assumptions, even though all assumptions used are
physically well-motivated. Can our classification of strongly
physical invariants for those levels be extended somehow into a
classification of physical invariants (the terms `physical' and `strongly
physical' are defined in Sec.1)? (ii) Can the methods developed here give
classification proofs for the other affine algebras?

A natural way to try to answer question (i) in the affirmative is to
apply Lemma 2 to weights other than just $\rho$, in other words to generalize
the proof of Lemma 4 to $\la'\ne \rho$.
There is a good chance this approach would work, but it could
result in a much lengthier argument.

The main thrust of our future research (see \eg \GR) will be directed towards
(ii), \ie applying
these arguments to other algebras, starting with the remaining rank 2
algebras and $A_1\oplus \cdots \oplus A_1$, and levels 2 and 3 of $A_n$. Sec.6
discusses our initial findings.

\medskip \noindent{NOTE TO READER}: Though the arguments contained in this
paper should be rigorously complete, this is only a preliminary version.
The final version will be co-authored with Patrick Roberts: he will try
to make it a little more accessible to physicists as well as include some
supplementary material (\eg exactly how the proof developed here relates
to the earlier $A_1$ classification proofs). This final version should be
completed, and submitted to the hep-th bulletin board, by early January
1993.

\bigskip
This work is supported in part by the Natural Sciences and Engineering
Research Council of Canada. I would particularly like to thank Patrick
Roberts for helping me understand parts of \MS, and for valuable assistance
in writing up this paper. I have also benefitted from conversations with
Quang Ho-Kim and C.S. Lam. I also appreciate the hospitality shown by the
Carleton mathematics department, where this paper was written.

\bigskip\bigskip\centerline{{\bf Appendix: $\rho$-coupling for $A_1$}}
\bigskip

An important step in the $A_2$ classification proof given in this paper
is the $\rho$-coupling Lemma proven in Sec.4. Its proof (see Claim 1 there)
assumes knowledge of the $\rho$-coupling lemma for $A_1$, given in
Sec.6. Because of this, and because the $\rho$-coupling proof for $A_1$ is
more transparent but similar in spirit to that of $A_2$, we have included
here the $A_1$ proof. After giving it, a few brief
comments on how to finish off a classification proof for $A_1$ are provided.
Claim 1 in Sec.4 is the $A_1$ $\rho$-coupling
Lemma, if we were to ignore the $A_1$ norm condition; its proof
will be completed at the end of this appendix.

The proof given below for $A_1$ $\rho$-coupling is certainly not intended to
be the shortest such; because our primary interest is in proving Claim 1,
we will exploit the $A_1$ norm condition $(A.1a)$ as rarely as possible.
\medskip

Write $k'=k+2=2^Lk''$, where $k''$ is odd. Define the integer $M$ by $k'/2
\le 2^M<k'$. Identify a weight $\la=m\beta_1$
of $\hat{A_1}$ by its Dynkin label $m$. Suppose $N_{1,a}>0$ for some $A_1$
positive invariant $N$ of level $k$. The norm condition reads
$$a^2\equiv 1\sp({\rm mod}\sp 4k').\eqno(A.1a)$$
Hence $a$ must be odd.
The parity $\epsilon(m)$ of some weight $m$ is simply
$$\epsilon(m)=\left\{\matrix{+1&{\rm if}\sp 0<\{m\}_{2k'}<k'\cr
-1&{\rm if}\sp k'<\{m\}_{2k'}<2k'\cr
0&{\rm if}\sp \{m\}_{2k'}=0\sp{\rm or}\sp k'\cr}\right. , \eqno(A.1b)$$
where throughout this Appendix we use the notation $\{x\}_y$ for the unique
number satisfying both $\{x\}_y\equiv x$ (mod $y)$ and $0\le \{x\}_y<y$.
Using this, Lemma 2 becomes
$$\eqalignno{0<\ell<k', &\sp \ell \sp{\rm relatively\sp prime\sp to\sp}2k',
\sp \Rightarrow\sp\{\ell a\}_{2k'}<k';&\cr
k'<\ell<2k', &\sp \ell \sp{\rm relatively\sp prime\sp to\sp}2k',
\sp \Rightarrow\sp\{\ell a\}_{2k'}>k';&(A.1c)\cr}$$
We want to find all integers $1\le a<k'$ satisfying both $(A.1a,c$).
Eqs.($A.1$) are the analogues of eqs.(4.1).

Assume first that $k'$ is odd (\ie that $L=0$), and define $N\ge 0$ so that
$a2^N<k'<a2^{N+1}<2k'$. If $a>1$, then $N<M$. Put $\ell=k'-2^{N+1}$; it will
lie between 0 and $k'$, and will be relatively prime to $2k'$. Then $(A.1c$)
implies $k'>\{ak'-a2^{N+1}\}_{2k'}=3k'-a2^{N+1}>k'$, a contradiction.
Therefore, $k'$ odd implies $a$ must equal 1.

Thus it suffices to consider $k'$ with $L>0$.
Let $a_2=\{a\}_{2^{L+1}}$. There are two different cases: either $a_2\le 2^L$
(to be called case 1), or $a_2>2^L$ (to be called case 2). If $k'=2^L$, there
will only be case 1. \medskip

\noindent{\it Consider case 1 first}.
Define $\ell_i=k''+2^i$, for $i=1,\ldots,M-1$. Then these $\ell_i$
will necessarily be relatively prime to $2k'$, and they all will lie in the
range $0<\ell_i<k'$. Let $b=a/k'$ and $c=1-a_2/2^L$. Then $(A.1c$) tells us
that no $1\le i<M$ can have $1\le {a_2\over 2^L}+\{2^ib\}_2\le 2$.
Write out the binary
expansion $b=\sum_{i=1}^\infty b_i 2^{-i}$ of $b$ (so each $b_i=0$ or 1).
Then we have, for each $i=1,\ldots,M-1$, that $\{ 2^i b\}_2> 1+c$ if $b_i=1$,
and $\{2^i b\}_2<c$ if $b_i=0$.

Assume inductively that $b_1=\cdots=b_n=0$ for some $1\le n<M-1$, but $b_{n+1}
=1$. Then $2^nb=\{2^nb\}_2<c$, but $2^{n+1}b=\{2^{n+1}b\}_2> 1+c$. Hence,
$1+c< 2^{n+1}b<2c$, \ie $1<c$, which is false.

A similar calculation holds if $b_1=\cdots=b_n=1$ but $b_{n+1}=0$. Thus there
are exactly two possibilities: either $b_i=0$ for all $i=1,\ldots,M-1$, or
$b_i=1$ for all $i=1,\ldots, M-1$ --- \ie either $a<k'/2^{M-1}$ or $a>k'-
k'/2^{M-1}$. But $k'/2^{M-1}\le 4$, so $a$ odd implies either $a=1$ or 3,
or $a=k'-1$ or $k'-3$. Eq.($A.1a$) now forces $a=1$ or (if $L=1$) $a=k+1$.
\medskip

\noindent{\it Case 2 is harder}, and we will begin by proving it for $L=1$.
As before, take $\ell_i=2^i+k'/2$ for $i=1,\ldots,M-1$. Eq.($A.1c$) however
now reads ${1\over 2}< \{2^ib\}_2<{3\over 2}$, since $a_2=3$ here.

Consider first $b_1=0$. Then ${1\over 2}\le 2b$ implies $b_2=1$, and $4b<
{3\over 2}$ implies $b_3=0$. In fact, $b_i$ continues alternating between
0 and 1, for $i=1,\ldots,M$. The same conclusion holds if $b_1=1$. Therefore,
for $M$ even, $a=k'/3+\epsilon$ or $a=2k'/3-\epsilon$, where $-k'/(3\cdot
2^M)\le\epsilon<k'/(3\cdot 2^{M-1})$, and for $M$ odd, $a=k'/3+\epsilon'$
or $a=2k'/3-\epsilon'$, where $-k'/(3\cdot 2^{M-1})\le \epsilon'<k'/(3\cdot
2^M)$. $\epsilon$ and $\epsilon'$ are fixed by the requirement that $a$ be
odd. There are 3 possibilities: if $k'\equiv 0$ (mod 3), we have $a=k'/3+1$
or $a=2k'/3-1$; if $k'\equiv \pm 1$, we have $a=k'/3\mp 1/3$.
Eq.($A.1a)$ tells us that for $k'\equiv 0$ (mod 3), these $a$ can only work
for $k'\equiv 30$ (mod 36); for $k'\equiv \pm 1$ (mod 3) they cannot
satisfy $(A.1c$).

So consider $k'\equiv 30$ (mod 36). Then $\ell=6+k'/6$ will be relatively
prime to $2k'$. We find that $\ell(k'/3+1)\equiv 6-k'/6$ (mod 2$k'$).
This then will contradict $(A.1c$), unless $6-k'/6>0$, \ie $k'<36$, \ie
$k'=30$. This concludes the proof of case 2, for $L=1$.

All that remains for us to prove is case 2 for $L>1$. Choosing $\ell_i=
2^i+k''$ gives us $c'< \{2^ib\}_2<1+c'$ for $i=1,\ldots,M-1$, where $c'=
2-a_2/2^L$. Choosing
$\ell_i'=k''-2^i$ gives us $1-c'<\{2^ib\}_2< 2-c'$ for $i=1,\ldots,M-L$.
Adding these, we get ${1\over 2}<\{2^ib\}_2<{3\over 2}$ for $i=1,\ldots,
M-L$. Therefore by the case 2 $L=1$ argument
we get that $b_i$ alternates between 0 and 1 for $i=1,\ldots,
M-L+1$. From this and the $\ell_i,\ell_i'$ inequalities we see that,
unless $M-L=1$, the binary expansion of $c'$ either looks like
$c'=0.10\ldots$ or $c'=0.01\ldots$ (in which case by the $\ell_i$
inequalities we cannot have $b_j=b_{j+1}=b_{j+2}$ for any $j<M$), and if
$M-L>2$ we have $c'=0.101\ldots$ or $c'=0.010\ldots$ (in which case we cannot
have $b_j=b_{j+1}$ for any $j<M$).

Now take $\ell_i''=2^i-k''$ for $i=M-L+1,\ldots,M$. This means either
$\{2^ib\}_2+c'<1$ or $\{2^ib\}_2+c'\ge 2$ for these $i$. Then by the case
1 argument, we must have $b_{M-L+1}=\cdots=b_M$.

Thus, either $M-L=1$, or both $M-L=2$ and $L=2$. If $M-L=2$ and $L=2$, then
$k'=4\cdot 5$ or $k'=4\cdot 7$. Otherwise $M-L=1$, \ie $k'=3\cdot 2^L$.
{}From the above
calculations we can read off that $a=k'/2\pm 1$ here. Eq.($A.1a)$ reduces to
$2^{L-2}\cdot 3\equiv \mp 1$ (mod 4). The  only possible solution is
$L=2$ (\ie $k'=12$) and $a=k'/2+1=7$.      \medskip

This completes the classification of the case 2 $\rho$-couplings $a$, and
hence the proof of the $\rho$-coupling lemma for $A_1$, except for 4 levels
where the argument broke down: $k=10,18,26,28$. These can be explicitly
worked out on a computer.

Little work now remains to obtain a new classification proof for $A_1$. The
$A_1$ permutation invariants can be easily enumerated using the
expression $S_{mn}^{(k)}=\sqrt{2/k'}\sin(\pi mn/k')$. Apart from the
exceptional level $k=10$, this classifies all physical invariants of level
$k\equiv 1,2,3$ (mod 4). To find the strongly physical invariants of level
$k\equiv 0$ (mod 4), the methods of Sec.5 suffice, and indeed reduce
ultimately to an example in \MS.

\medskip\noindent{\it Proof of Claim 1 in Sec.4}\quad If we replace the
$K$ in Claim 1 with $k'=k+2$ here, we see that it is simply the $A_1$
situation,
ignoring the norm condition. The only places we used $(A.1a)$ were in the
case 1 proof when we eliminated
$a=3$ etc.; the case 2 proof for $L=1$, when we eliminated all but $k'\equiv
30$
(mod 36); and the case 2 proof for $L>1$, when we threw out $k'=2^L\cdot 3$
for $L>2$.

Note that $a$ will satisfy $(A.1c$) iff $k'-a$ will, so it suffices to
consider $a\le k'/2$.

Consider first the case 1 proof, and $a=3$. Because $3=a_2<2^L$, we must
have $L>1$. If $k'\equiv -1$ (mod 3) use $\ell=(k'+1)/3$, while if
$k'\equiv +1$ (mod 3) use $\ell=(k'+2)/3+k''$. If $k'\equiv 0,3$ (mod 9)
take $\ell=k'/3+1$, while if $k'\equiv -3$ (mod 9) use $\ell=k'/3+3$ (this
fails for $k'<9$, but there are no such $k'$ divisible by 4 and 3).

Now consider the case 2 proof for $L=1$.
Taking $\ell=3$ eliminates $k'\equiv -1$ (mod 3), and
for $k'>12$ taking $\ell=k'/2+6$ eliminates $k'\equiv +1$ (mod 3). The only
$k'\le
12$ with $k'\equiv +1$ (mod 3) and $L=1$ is $k=8$. For $k'\equiv 6$ (mod
36) use $\ell=4+k'/6$ (this fails for $k=4$), and for $k'\equiv 18$ (mod 36)
use $\ell=2+k'/6$.

Finally, consider the case 2 proof for $L>1$, where $k'=2^L\cdot 3$. For
$k'>14$, take $\ell=7$ (this fails for $k=10$).

The only $k=K-2$ which escaped our arguments are $k=4,8,10,18,26,28$. These can
be explicitly worked out. This concludes the proof of Claim 1.

\bigskip\bigskip \centerline{{\bf References}} \bigskip

\item{\WN} E. Witten, {\it Commun. Math. Phys.} {\bf 92} (1984) 455;

\item{} S.P. Novikov, {\it Usp. Mat. Nauk} {\bf 37} (1982) 3

\item{\GW} D. Gepner and E. Witten, {\it Nucl. Phys.} {\bf B278} (1986) 493

\item{\HAL} V.G. Ka{\v c}, {\it Funct. Anal. App.} {\bf 1} (1967) 328;

\item{} R.V. Moody, {\it Bull. Am. Math. Soc.} {\bf 73} (1967) 217;

\item{} K. Bardakci and M.B. Halpern, {\it Phys. Rev.} {\bf D3} (1971)
2493

\item{\KAC} V.G. Ka{\v c}, {\it Infinite Dimensional Lie Algebras}, 3rd ed.,
(Cambridge University Press, Cambridge, 1990)

\item{\CIZ} A. Cappelli, C. Itzykson and  J.-B. Zuber,
{\it Nucl. Phys.} {\bf B280 [FS18]} (1987) 445;

\item{} A. Cappelli, C. Itzykson and  J.-B. Zuber,
{\it Commun. Math. Phys.} {\bf 113} (1987) 1;

\item{} A. Kato, {\it Mod. Phys. Lett.} {\bf A2} (1987) 585;

\item{} D. Gepner and Z. Qui, {\it Nucl. Phys.} {\bf B285} (1987) 423

\item{\ITZ} C. Itzykson, {\it Nucl. Phys. (Proc. Suppl.)} {\bf 5B}
(1988) 150;

\item{} P. Degiovanni, {\it Commun. Math. Phys.} {\bf 127} (1990) 71

\item{\COMM} T. Gannon, {\it ``WZW Commutants, Lattices, and Level 1
Partition Functions''} (Carleton preprint, 1992)

\item{\RTW} Ph. Ruelle, E. Thiran and J. Weyers,  {\it
Commun. Math. Phys.} {\bf 133} (1990) 305

\item{\HET} T. Gannon, {\it ``Partition functions for heterotic WZW
 conformal field theories''} (Carleton preprint, 1992)

\item{\CR} P. Christe and F. Ravanani, {\it Int. J. Mod. Phys.} {\bf
A4} (1989) 897

\item{\MS} G. Moore and N. Seiberg, {\it Nucl. Phys.} {\bf B313} (1989) 16

\item{\HK} Q. Ho-Kim and T. Gannon, {\it ``The low level modular invariant
partition functions of rank 2 algebras''} (work in progress)

\item{\KMPS} P. Goddard and D. Olive, {\it Int. J. Mod. Phys.} {\bf A1}
(1986) 303;

\item{} S. Kass, R.V. Moody, J. Patera and R. Slansky, {\it Affine Lie
Algebras, Weight Multiplicities, and Branching Rules} Vol.1 (University
of California Press, Berkeley, 1990)

\item{\CS} J.H. Conway and N.J.A. Sloane, {\it Sphere packings,
Lattices and Groups}, (Springer-Verlag, New York, 1988)

\item{\RT} P. Roberts and H. Terao,
{\it Int. J. Mod. Phys.} {\bf A7} (1992) 2207;

\item{} N.P. Warner, {\it Commun. Math. Phys.} {\bf 130} (1990) 205

\item{\SCH} B. Gato-Rivera and A.N. Schellekens, {\it Nucl. Phys.} {\bf
B353} (1991) 519

\item{\BI} M. Bauer and C. Itzykson, {\it Commun. Math. Phys.} {\bf 127}
(1990) 617

\item{\NM} O. Perron, {\it Math. Ann.} {\bf 64} (1907) 248;

\item{} G. Frobenius, {\it S.-B. K. Preuss. Akac. Wiss. Berlin} (1908) 471;
(1909) 514; (1912) 456

\item{\PM} H. Minc, {\it Nonnegative Matrices}, (John Wiley \& Sons,
New York, 1988);

\item{} E. Seneta, {\it Non-Negative Matrices}, (George Allen \& Unwin Ltd,
London, 1973)

\item{\ALZ} D. Altsch\"uler, J. Lacki and Ph. Zaugg, {\it Phys. Lett.}
{\bf B205} (1988) 281

\item{\VS} D. Verstegen, {\it Nucl. Phys.} {\bf B346} (1990) 349

\item{\VER} E. Verlinde, {\it Nucl. Phys.} {\bf B300 [FS22]} (1988) 360;

\item{} G. Moore and N. Seiberg, {\it Phys. Lett.} {\bf B212} (1988) 451

\item{\BMW} L. B\'egin, P. Mathieu and M. Walton, ``$\hat{su}${\it (3)
fusion coefficients''} (Laval preprint PHY-22, 1992)

\item{\ROB} P. Roberts, {\it Whatever Goes Around Comes Around:
Modular Invariance in String Theory and Conformal Field Theory}, Ph.D. Thesis
(Institute of Theoretical Physics, G\"oteborg, 1992)

\item{\GR} T. Gannon, {\it ``Towards a classification of
SU(2)$\oplus\cdots\oplus $SU(2) modular invariant partition functions''}
(work in progress)

\item{\VT} D. Verstegen, {\it Commun. Math. Phys.} {\bf 137} (1991) 567

\item{\WAL} M. Walton, {\it Nucl. Phys.} {\bf B322} (1989) 775;

\item{} D. Altsch\"uler, M. Bauer and C. Itzykson, {\it Commun. Math. Phys.}
{\bf 132} (1990) 349

\end